\documentclass{aastex701}
\usepackage{url}
\usepackage{siunitx}
\usepackage{upgreek}
\usepackage{booktabs}
\usepackage{multirow}

\begin{document}

\title{A Slit Mask Integral Field Unit for the Robert Stobie Spectrograph on the Southern African Large Telescope: I. Instrument Development}

\author[0000-0003-1601-8048,gname='Sabyasachi',sname='Chattopadhyay']{Sabyasachi Chattopadhyay}
\affiliation{Space Science and Engineering Initiative, College of Engineering, University of Hawai’i at Manoa, 2540 Dole Street, Honolulu, HI 96822, USA}
\affiliation{Centre for Space Research, North-West University, Potchefstroom Campus,
Private Bag X6001, Potchefstroom, 2520, South Africa.}  
\affiliation{South African Astronomical Observatory, 1 Observatory Rd, Observatory, Cape Town, 7925, South Africa}
\email[show]{sabya@hawaii.edu}

\author[0000-0002-3131-4374, gname='Matthew A.', sname='Bershady']{Matthew A. Bershady}
\affiliation{University of Wisconsin, Department of Astronomy, 475 North Charter Street, Madison, WI 53706, USA}
\email[hide]{mab@astro.wisc.edu}

\author{Michael P. Smith}
\affiliation{University of Wisconsin, Department of Astronomy, 475 North Charter Street, Madison, WI 53706, USA}
\email[hide]{mps@astro.wisc.edu}

\author[0000-0002-6518-781X]{Antoine Mahoro}
\affiliation{South African Astronomical Observatory, 1 Observatory Rd, Observatory, Cape Town, 7925, South Africa}
\email[hide]{antoine@saao.ac.za}

\author[0000-0002-7827-6184]{Marsha J. Wolf}
\affiliation{University of Wisconsin, Department of Astronomy, 475 North Charter Street, Madison, WI 53706, USA}
\email[hide]{mwolf@astro.wisc.edu}

\begin{abstract}
Integral field spectroscopy (IFS) has been added as a new observation mode to the Robert Stobie Spectrograph (RSS), the workhorse multi-mode instrument on the Southern African Large Telescope. RSS operates as an imaging spectrograph covering 320-900 nm with a spectral resolution--slit-width product of 6600 arcsec. Using fiber optics and prismatic fold mirrors, we have been able to construct compact integral field units (IFUs) that fit within the same volume as the long-slit cassettes (134 mm × 130 mm × 8 mm). These `slit mask' IFUs (SMIs) direct the telescope beam into a 2D sky-facing fiber array routed in the focal plane dimension into an 8$\arcmin$ 1D pseudo-slit, with fiber output redirected back into the spectrograph collimator. The first completed unit, SMI-200, features 303 object fibers and 24 sky fibers, providing a spatial resolution of 0.8$\arcsec$ (200  $\upmu$m core diameter) over a field of view (FOV) of 22.5$\arcsec$ × 17.6$\arcsec$. This paper describes the specific design considerations and design and fabrication strategies to maximize performance and minimize risk during construction, given the demanding and highly constrained cassette geometry. We also detail mapping and laboratory characterization of the IFU. Laboratory measurements demonstrate a total throughput of  77\%, but an effective throughput of only 55-60\% within the RSS collimator acceptance beam of f/4.2 due to losses dominated by focal ratio degradation (FRD) induced by the detailed routing within the tight cassette volume.

\end{abstract}
\noindent
\keywords{Integral Field Spectroscopy; Fiber-fed IFU; Retrofit instrument}

\section{Introduction}
\label{sec:intro}

With the exception of the Southern African Large Telescope \citep[SALT;][]{stobie2000}, all 8m-class telescopes today offer seeing-limited integral field spectroscopy (IFS) well suited for nebular spectroscopy at optical wavelengths. The most dramatic are MUSE on the VLT \citep{muse}, VIRUS on the HET \citep{Hill2021}, and LLAMAS on Magellan \citep{Furesz2020}. Although using different coupling technologies (image slicer, fibers, and lenslet-coupled fibers respectively), these instruments have enormous spatial multiplex by virtue of replicated spectrographs. However, they are limited to fixed spectral configurations. Other IFS on 8m-class telescopes, such as MEGARA on the GTC \citep{megara} and KCWI on Keck \citep{Morrissey2018}, have powerful single spectrographs (in the case of KCWI with two spectral channels) that allow for a variety of spectral configurations that optimize wavelength coverage and spectral resolution for a given scientific program. While all the above have been purpose-built for IFS, several other imaging spectrographs, originally designed as long-slit or multi-slit systems, have been augmented to have IFS capabilities. These include GMOS on Gemini \citep{Allington2002}), ESI on Keck \citep{Sheinis2006}, FOCAS on Subaru \citep{Ozaki2020}, and IMACS and M2FS on Magellan \citep{McGurk2020,Mateo2022}. These uses lenslet-coupled fibers or image slicers. However, SALT does have  an imaging spectrograph, RSS \citep{rss_opt,rss_ops,rss_mech}, located at the top, corrected imaging port of the prime focus assembly. Like GMOS, RSS is suited to the implementation of a fiber-based IFS augmentation. To date, RSS serves primarily as an interchangeable slit-based instrument that spans 320-900~nm over varying spectral bands and resolution through the choice of Littrow VPH gratings and camera articulation angles \citep{rss_ops}. Although RSS was implemented with a Fabry-Pérot mode for imaging spectroscopy via insertable etalon pairs \citep{Rangwala2008}, the performance and calibration of this mode have suffered from non-uniform coating reflectivity and ghosting \citep{Williams2016}, coupled with the variable SALT pupil and the absence of simultaneous monitoring for precision relative flux calibration; this mode has not been available since 2019 \citep{Romero-Colmenero2020}. Slit-based imaging spectroscopy is accomplished via a suite of varying-width apertures for up to 8$\arcmin$ longslit or multi-slit configurations, each residing in cassettes selectable via a mechanized magazine. The addition of IFU capabilities for RSS takes advantage of these cassettes, as described below, reopening imaging spectroscopy with new scientific capabilities for SALT.  
 
Imaging spectrographs have often been preferred over fiber-fed spectrographs for preserving spatial information and maximizing throughput, particularly when large patrol fields for multi-object spectroscopy are not needed. Early attempts at spectrograph fiber coupling often ran afoul of focal-ratio mismatch between telescope and spectrograph due to unaccounted for focal-ratio degradation (FRD) from poorly terminated fibers often fed with unsuitably slow beams. With proper design and termination, FRD losses can be minimized. Similarly, surface losses can be further reduced with modern anti-reflection coatings \citep{drory}. These developments make fiber-based spectrograph coupling competitive with imaging spectrographs directly coupled to the telescope focal plane. For telescopes such as SALT, with inherent changes in the telescope pupil over the course of an observation, fiber-coupled spectroscopy becomes a strong advantage. This is because of the far-field azimuthal scrambling properties of circular-core fibers. This scrambling ensures that internal spectrograph illumination, particularly of the dispersing elements, remains as uniform as possible. This is critical for calibration and sky-subtraction. Radial scrambling of the near-field, particularly with non-circular fibers, has been a boon for precision radial velocity measurements, where the spatial information due to changes in slit illumination (due to tracking errors or changes in the point-spread function) degrades stability. In principle, by transposing the telescope pupil into the near-field, e.g., with lenslets, pupil illumination can be made even more uniform than simply direct illumination onto circular fiber cores. Lenslet coupling is beyond the scope of what can be achieved in the SMI mechanical envelope described below. This is because efficient lenslet coupling \citep{chattopadhyay_2022_JATIS} requires a much slower beam than provided by SALT, and there is insufficient room to provide a focal expander, as done, e.g., for the GMOS IFU \citep{Allington2002}. Nonetheless, azimuthal scrambling of the far field with circular fibers alone offers a significant advantage, and led to the initial concept of the pupil-scrambling IFU for SALT described in \citep{Smith2016}, which forms the basis of the SMI program.

\begin{figure}[ht]
     \centering
     \includegraphics[width=0.9\linewidth]{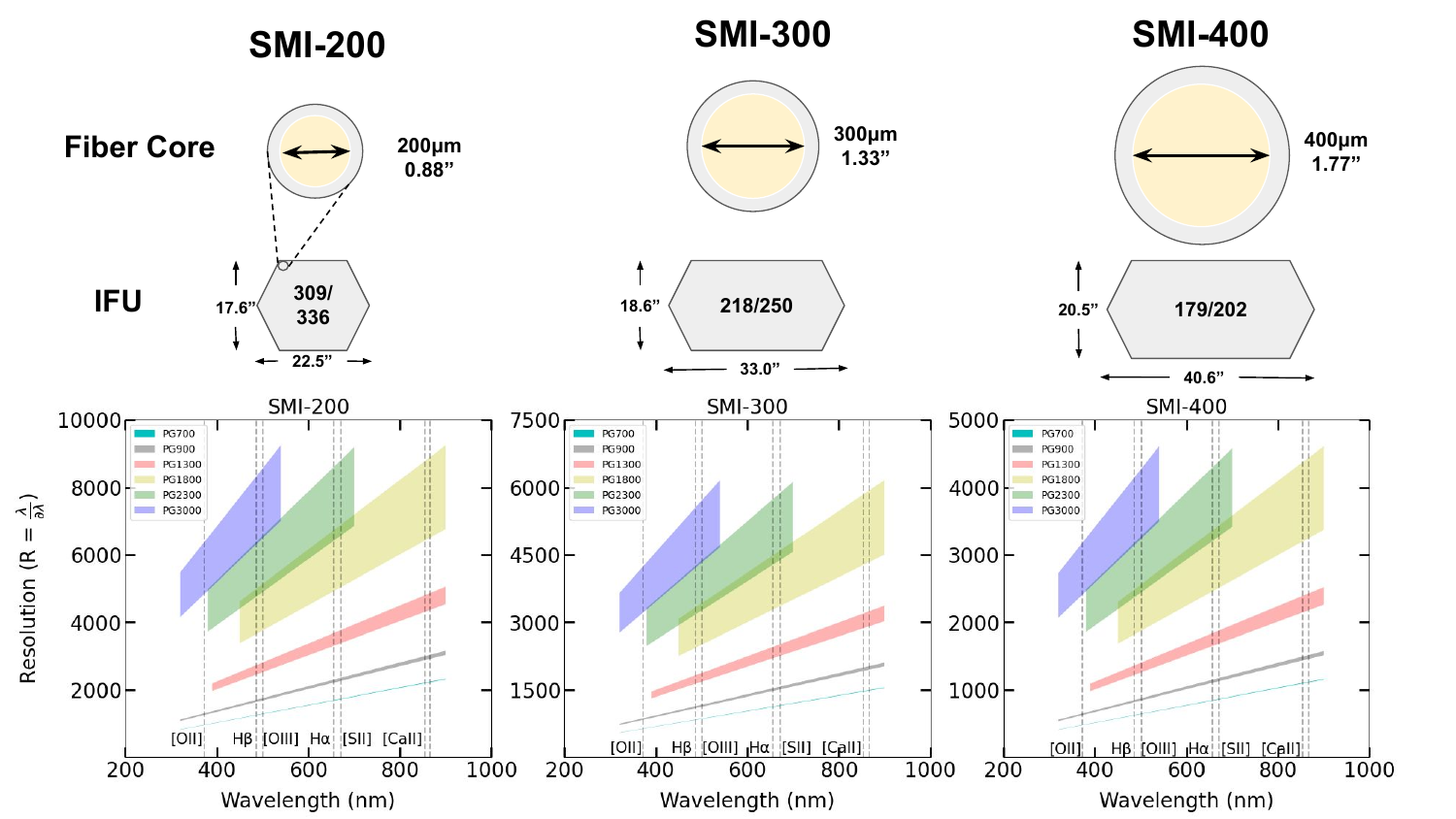}
     \caption{Different SMI variations to serve different spatial coverage, sampling, and spectral resolution. Top row: effective fiber sizes. Middle row: effective IFU footprint. Numbers inside the IFUs give the ideal IFU and total fiber count (including sky) for each SMI. In the case of SMI-200, the usable number of fibers is 327 (303 for the object IFU and 24 for sky). Bottom row: Spectral resolution as a function of wavelength delivered by RSS using the six available Littrow VPH gratings (PG700, PG900, PG1300, PG1800, PG2300, and PG3000, where the number refers to the line-density in mm$^{-1}$) with different SMIs (200, 300, 400, left to right).
     }
     \label{fig:smi}
\end{figure}

The SAAO Astrophotonics Lab is developing a suite of circular-fiber IFUs with varying spatial sampling (0.88$\arcsec$, 1.33$\arcsec$, 1.77$\arcsec$) and on-sky footprint (22.5$\arcsec$$\times$17.6$\arcsec$, 33$\arcsec$$\times$18.6$\arcsec$, 40.7$\arcsec$$\times$20.5$\arcsec$) using different core diameters (200 $\upmu$m, 300 $\upmu$m, 400 $\upmu$m) fibers.\footnote{We adopt the prime focus plate scale of 226.178~$\upmu$m/$\arcsec$.} These are illustrated in Figure~\ref{fig:smi}. The IFUs are constructed within the volume (130 mm$\times$135~mm$\times$8~mm) of slit mask cassettes, and hence they are called Slit Mask IFUs (SMI) \citep{chattopadhyay2022,chattopadhyay2024}. This paper describes the first SMI of this series, the SMI-200, which uses 200~$\upmu$m core diameter fibers, which is now in routine operations. In the following section, we describe the scientific opportunities as well as optical and opto-mechanical constraints implementing the SMI as a retrofit to the existing RSS instrument (\S~\ref{sec:considerations}). Details of the SMI design and fabrication methods are found, respectively, in \S~\ref{design} and \S~\ref{sec:assembly}. We provide descriptions of the lab calibration machine (SWiFT) as well as the measurement procedure and fiber performances in \S~\ref{sec:stray} and in the Appendices. Finally, we discuss lessons learned from the initial SMI-200 fabrication, and compare the SMI series instruments with comparable IFS on other 10~m-class telescopes in \S~\ref{sec:sum}. In a separate paper (Paper II, in preparation) we  describe the observing strategy particular to SMI-200 on SALT – a fixed-elevation telescope with limited acquisition and guiding precision; mechanical calibration of on-telescope performance (insertion repeatability, focus, field alignment, trace contrast, instrument throughput and spectral resolution); the data reduction pipeline (aperture extraction, wavelength and relative flux calibration, sky subtraction, cube generation); and science verification (comparison of kinematic, flux, and line-ratio maps to measurements in the literature).

\section{Instrument design considerations}
\label{sec:considerations}

\subsection{Robert Stobie Spectrograph: scientific opportunities for IFS}
\label{sec:opportunites}

RSS is a formidable single spectrograph for IFS given its large collimated beam pupil (150~mm diameter) coupled to an 8$\arcmin$ field of view on telescope with an 11~m pupil, a comprehensive VPH grating complement achieving a resolution $\times$ slit-width product (R$\phi$) up to 6600 arcsec over most of its wavelength range, and a large, low-noise ($\sim$2e$^-$ rms) high quantum-efficiency ($>$80\% between 350-600~nm) detector focal-plane (6144 $\times$ 4096 pixels) sampling a 16 deg, f/2.2 camera. The spectrograph optics and detector well-sample entrance apertures as small as 0.6$\arcsec$; the detector scale is 0.1267 arcsec per unbinned pixel. However, for spatial mapping and sky-subtraction, there are major advantages for IFS  over long-slit or multi-slit modes, as we proceed to motivate. These advantages offer a wide variety of compelling science applications that were previously prohibitive on SALT. As we also show, SMI IFS is also highly competitive compared to FP when the science requires a premium on spectral coverage, as is so often the case for stellar kinematics as well as stellar chemical and gas-phase diagnostics.

\subsubsection{Mapping efficiency of extended objects} 

SMI IFS offers a significant advantage over RSS long-slit spectroscopy for extended sources, e.g., clusters, galaxies, or Galactic nebulae with angular diameters comparable to the slit-width $\times$ slit-length product of RSS. For the 8' slit-length and a seeing-match slit-width of 1.5$\arcsec$, the characteristic diameter is roughly 30$\arcsec$. This characteristic size scales as the square root of the effective slit-width, where the slit-width is driven by the scientific needs for spatial and spectral resolution. 

\begin{figure}[ht]
    \centering
    \includegraphics[width=0.85\linewidth]{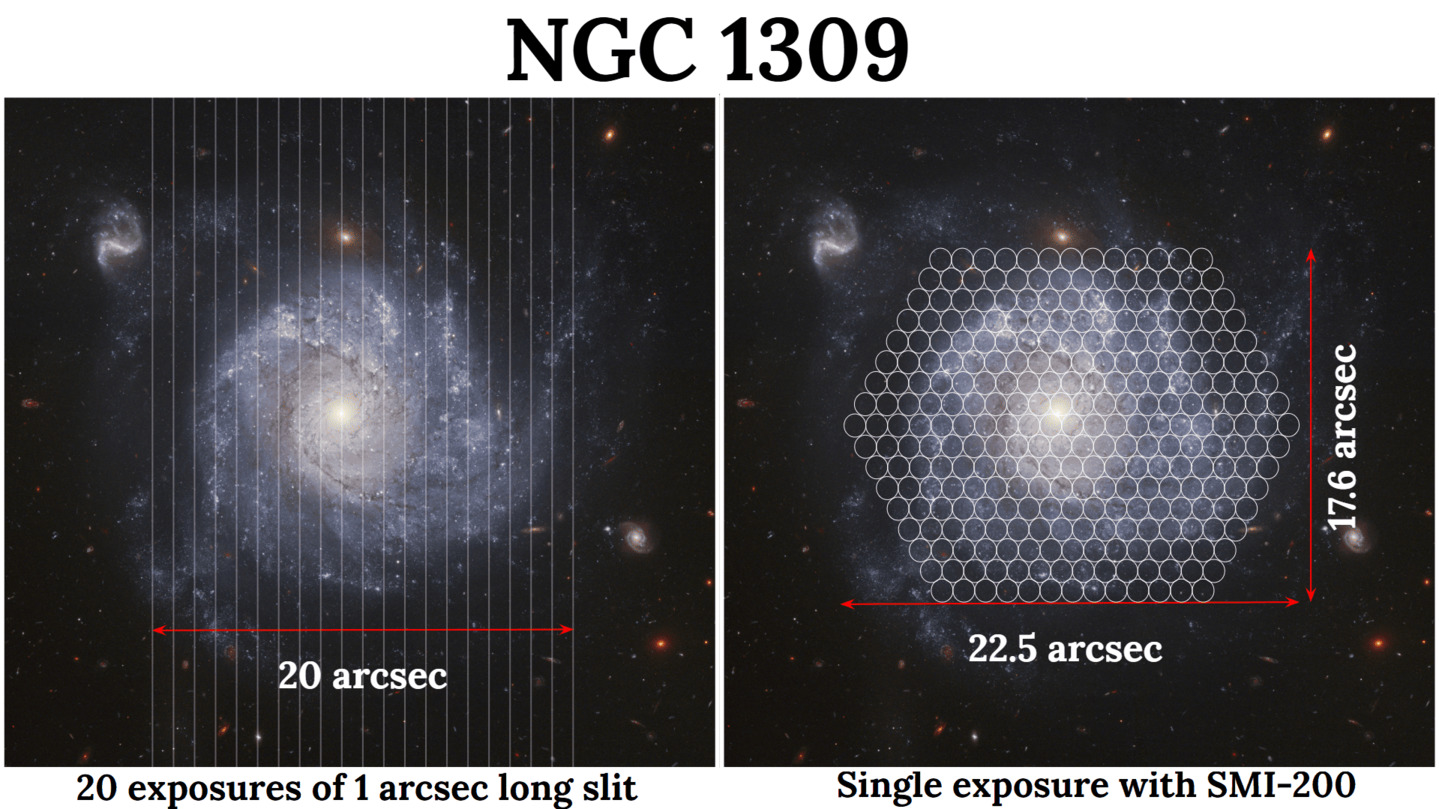}
    \caption{The observational efficiency advantage of using SMI-200 over a 1 arcsec wide longslit. SMI-200 can observe objects (such as the NGC 1309, at $z\sim0.007$, shown here) with $\sim$20-30$\arcsec$ diameter in a single exposure (right panel), which would take $\sim$20 separate exposures to be covered by a 1 arcsec longslit (left panel) at slightly lower spatial and spectral resolution. The optical image (F435W, F555W, and F814W bands) of NGC1309 is obtained from the Hubble archive (NASA, ESA, STScI/AURA).}
    \label{fig:ls2ifu}
\end{figure}

For SMI-200, with 0.88$\arcsec$ fiber sampling diameter ($D$) and an effective slit-width $\phi=\cos(\pi/6)D=0.76\arcsec$ (roughly 6 unbinned pixels), the characteristic size is roughly 20$\arcsec$, with spectral resolutions up to $\sim$8700 achievable. The spectral resolving power corresponds to a velocity dispersion of $\sim$15~km/s, typical of the ionized gas in H~II regions and stellar populations in the outer parts of galaxy disks. Figure~\ref{fig:ls2ifu} gives an illustrative comparison for mapping a nearby face-on galaxy using a 1$\arcsec$ long-slit. While the comparison is imperfect (SMI-200 achieves 25\% higher spatial and spectral resolution), it shows that SMI-200 can map this galaxy at a greater than 20 times faster rate for a given exposure time. Even accounting for throughput losses due to the fiber coupling (see Paper II), SMI-200 remains more than 10 times faster for mapping to a given depth (signal-to-noise), effective to long-slit measurements. This is particularly relevant for SALT, a queue-scheduled telescope with fixed elevation, since more than several visits to single targets in a given semester is rarely achieved. As shown in Figure~\ref{fig:smi}, the SMI series provides optimal coverage of extended sources over a range of apparent size (20 to 40$\arcsec$) and elongation ($0.5<\rm b/a<0.8$). This is well suited for observing galaxy clusters and groups, nearby individual galaxies over a wide range of inclination, merging and interacting systems, Local Group dwarf galaxies, and Milky Way H~II regions.

\subsubsection{Sky subtraction advantages} 
\label{sec:ss}

\begin{figure}[ht]
    \centering
    \includegraphics[width=\linewidth]{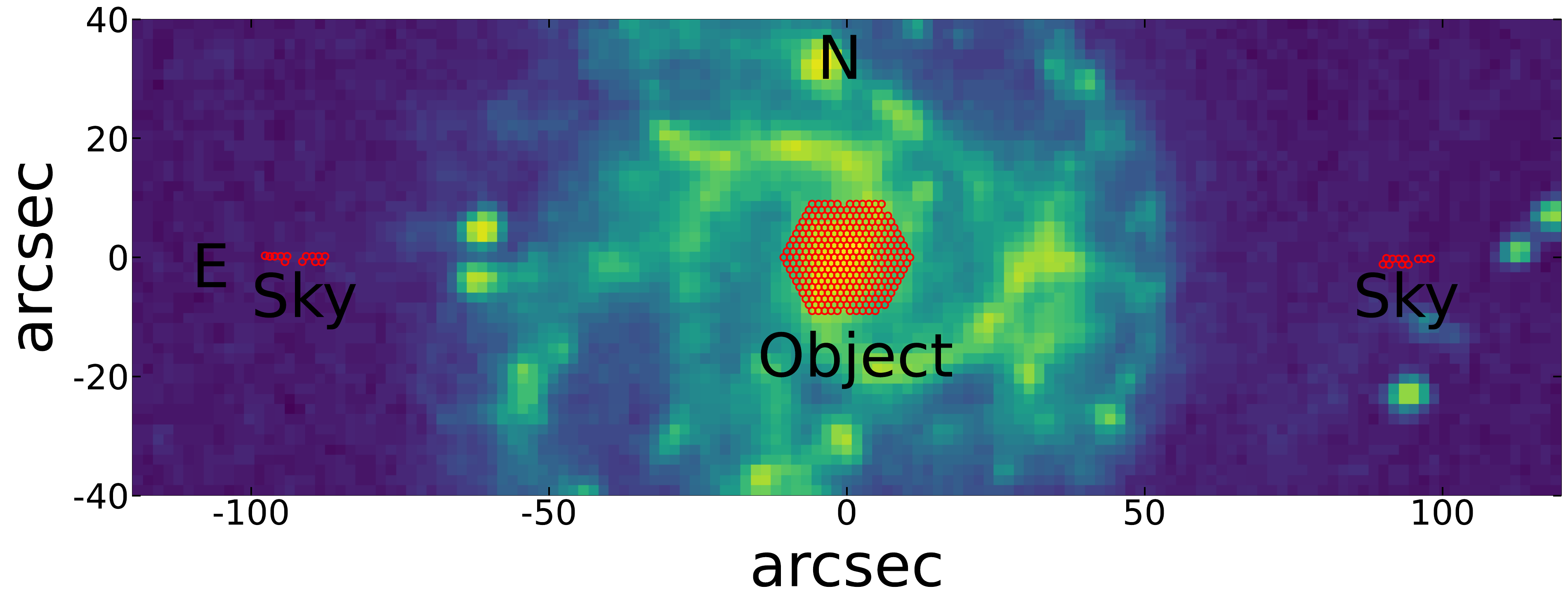}
    \caption{SMI-200 object (center) sampling array and sky (on the two sides) sampling array footprint over a blue DSS image of NGC 5468, a nearly face-on spiral galaxy at $z\sim0.01$. Each red circle denotes a 0.88$\arcsec$ diameter SMI-200 fiber core.}
    \label{fig:fp}
\end{figure}

Accurate removal of sky foregrounds in ground-based astronomical data requires both good temporal sampling of the intrinsic variation of the emission-line intensities and spatial sampling of the spectrograph line-spread function (LSF). Imaging slit spectroscopy for compact or stellar sources solves this naturally by providing sky apertures adjacent to the source on the sky and on the slit. For extended sources, this can be accomplished via nod-and-shuffle techniques \citep{Cuillandre1994}, but at the cost of observing efficiency. Multi-fiber spectroscopic systems, if properly designed, offer a more efficient means to simultaneously sample sky and the spectrograph LSF by interleaving an optimum number of sky fibers with object fibers along the pseudo-slit. As illustrated in Figure~\ref{fig:fp}\footnote{This figure uses a Digitized Sky Survey (DSS) image based on POSS-II plates, produced at the Space Telescope Science Institute under U.S. Government grant NAG W-2166.} for NGC 5468, this can be particularly advantageous for resolved study of the central regions of nearby galaxies, e.g., circumnuclear star-forming rings, counter-rotating core kinematics, and the interplay between AGN and the host ISM and stellar populations. For the SMI design, we take advantage of the 8$\arcmin$ patrol field by placing sky fibers at 90$\arcsec$ distance from the object IFU on both sides. These fibers are interleaved with the object IFU fibers. This interleaving and the detailed trades of the sky fiber number and location are described below.

However, there is a subtle but critical issue for sky subtraction due to the impact of SALT's moving pupil has on the spectrograph response. For a fixed-altitude telescope such as SALT, the primary mirror illumination seen at prime focus is a function of the hour angle of observation; during a given primary payload ``track" of an object, the primary mirror illumination changes with time. For RSS, when used as an imaging spectrograph (with slits or FP), this change manifests as a change in the internal illumination of the spectrograph optics, most significantly at the internal pupil where the dispersing elements are located. The pupil illumination is both a function of time and field position. The VPH gratings have known non-uniformities across their clear aperture, such that in addition to the time-varying emission-line intensities from the sky, the spectrograph response changes with time, wavelength, and field position. This has been observed to lead to emission line strengths varying with position along the slit that varies in wavelength and time \citep{Smith_2016}. Such behavior presents radical challenges for accurate sky subtraction.

In this case, fibers serve as a scrambling agent of the partially filled, time and position-varying light cone delivered by the telescope at the spectrograph input focal plane. With SMI fibers providing scrambling of the output near and far field, the grating plane becomes more uniformly illuminated and less variable with time and field position \citep{Smith_2016}. We anticipate this scrambling, coupled with good slit sampling of the sky fibers, to aid significantly in improving the accuracy of sky subtraction.

One of the reasons why accurate sky subtraction is important for SALT is to take full advantage of the large grasp realized by the large telescope and spectrograph pupils, thereby enabling RSS to probe low surface-brightness regions of Galactic nebulae, galaxies, and clusters. While SMI-200 fibers' $0.88\arcsec$ diameters  allow it to sample, e.g.,$\sim$10~pc for the Magellanic Clouds and 3.5~kpc for galaxies at z$\sim$0.02, SMI can also be used as a light-bucket \cite[e.g.,][]{Weijmans_2009}. With the light-bucket approach, the SMI-200 footprint matches very closely with, e.g., nearby Fornax cluster dwarfs. Similarly, SMI-200 can prove to be ideal for the observation of diffuse ionized gas in the Milky Way and external galaxies.

\subsubsection{Comparison to Fabry-Pérot imaging}
\label{sec:fpcomp}

Finally, here we make a comparison of the IFS offered by the SMI series to the RSS FP mode. The wavelength-dependent spectral resolution using SMI as the front-end instrument is shown in Figure \ref{fig:smi} for the six VPH gratings used to cover the instrument’s full resolution-wavelength range. It is relevant, in this context, to compare the resolution area product of the SMI to the FP system \citep{Rangwala2008}. RSS has a 150 mm internal pupil in a collimated beam, located near the dispersing elements, which corresponds to the Fabry-Perot etalons' clear aperture. This number can be used to compute the size of the bull's-eye (Jaquinot spot) within which $\lambda/\delta\lambda < R$, where $\delta\lambda$ is the change in the central transmitted wavelength through the etalon as a function of beam angle. The bull's eye is given by the beam angle on the etalon $\alpha=\sqrt(2/R)$. The beam angle on the etalon ($\alpha$) is related to the angle on the sky ($\theta$) by $\theta=\alpha \rm D_e/D_T$, where $\rm D_e$ and $\rm D_T$ are the etalon and telescope pupil diameters (150~mm and 11~m respectively), yielding $\theta \sim 56 (R/5000)^{-1/2}$ arcsec. Twice this angle can be compared to the angular extent of the SMI illustrated in Figure \ref{fig:smi}. While the FP at comparable spectral resolution still would provide significantly larger areal coverage than SMI, they would not be capable of simultaneously capturing the significant wavelength coverage afforded by the fiber-based pseudo-slit of the SMI. Hence, SMI would remain complementary should the FPs become operational again. The advantage of slit-based IFS is particularly strong for programs where wavelength coverage of extended sources is a scientific premium. This is often the case for spectral diagnostics (e.g., line-ratios for nebular emission, and abundance measurements for stellar absorption), as well as for stellar kinematic measurements that utilize many weak, intrinsically narrow lines.

\subsection{Robert Stobie Spectrograph and SALT: opto-mechanical constraints}
\label{sec:constraints}

The most significant constraints that arise from RSS are the combination of a physically tight, non-telecentric focal plane fed at f/4.2, for an instrument whose general functionality as a long-slit and multi-slit spectrograph must be preserved during routine use. Since the RSS collimator is designed for f/4.2, the beam speed is somewhat too slow to avoid focal ratio degradation (FRD) losses altogether in the fiber coupling without the addition of reimaging optics. However, RSS sits at the prime focus of SALT, with the first (field) lens sitting 10 mm above the focal plane, and the RSS guider 6~mm below the focal plane. There is also little lateral volume to utilize (e.g., via a macro relay) near the focal plane. In short, there is essentially no opportunity to augment the beam speed prior to fiber coupling. This is highly constraining for IFS implementation, particularly since any augmentation must be removable during routine operations.

The solution devised originally by \cite{Smith_2016} is to accept the telescope input beam as is, and take advantage of the cassette mechanisms used to house and insert the long-slit masks into the focal plane. The challenge lies in fitting a complex system of small fold mirrors and fibers near the focal plane into the volume of one cassette. This volume is 130 mm$\times$135 mm$\times$8 mm. At the same time, it is necessary to ensure the cassette outer structure is compatible with the elevator system that stores existing cassettes in a rack, and brings them to the focal plane upon demand. 

The volume constraints are critical in all dimensions. The smallest dimension (8 mm) is along the optical axis of the prime focus converging beam. This cassette thickness prevents us from implementing micro-lens coupling of fibers either to transpose the pupil or reduce the interstitial gap between fiber cores. (For the latter, instead, a dithering strategy at the telescope has been implemented to obtain integral field coverage.) The cassette thickness also fundamentally limits the size of any IFU in one dimension. The constraint from the mechanical envelope is $<$9~mm, and involves consideration of the fold-prism mechanical and clear aperture sizes, prism material refractive index, and their effective position within the cassette, described below. The manifestation of the constraint for the SMI series is the similar height of the units, and a natural elongation of the IFUs for larger fiber sizes to take advantage of the additional grasp. The natural elongation also suits the SMI series of IFUs to efficiently cover partially inclined galaxies.

In the lateral dimensions, the cassette size requires fibers to bend tightly. This is exacerbated by wanting the IFU near the field center where the telescope vignetting and non-telecentricity (NT) are minimized, yielding bend diameters of roughly 1/2 of the cassette dimension ($\sim$65~mm). Central positioning also minimizes fiber overlap within the 8~mm vertical volume during routing between the IFU and pseudo-slit. Even if this constraint were relaxed, such small bend radii are inevitable when routing fibers to populate the center of the fiber pseudo-slit. We undertook laboratory measurements of single-fiber FRD, and found that the FRD losses were only about 10\% for 200~$\upmu$m core fiber fed at f/4.2, and little (2 to 3\%) modulated by the induced bend-diameter. Similar measurements for 400$\mu$m core fiber yielded 5 to 6\% additional FRD-induced losses when inducing the bend-diameter. This gave us optimism that we could achieve high-throughput devices using fiber with core sizes in the 200 to 400$\mu$m range. As we see below for SMI-200, we did not account for the mechanical stress from bending many fibers bundled together, the impact of which is yet to be determined for large-fiber devices.

Another significant constraint stems from  SALT having a flat but highly non-telecentric focal plane, with the telecentric angle at any field position varying by $\sim$76 times the on-sky radial distance from the field center. The RSS collimator is designed to accept this NT. Thus, the Slit Maks IFUs need to be designed to capture and deliver the same optical prescription. While an effectively small IFU near the field center subtends only a small range of non-telecentric input beam angles, the output pseudo-slit spans the full 8$\arcmin$ field. This requires the fibers to be suitably splayed as a function of field position. At the end of the slit (a field position of 4$\arcmin$), the telecentric beam is tilted by roughly 5 deg from the focal-plane normal. 

Finally, the RSS optics \citep{rss_opt} are complicated by an all-refractive design constrained to perform over 320-900~nm and (particularly) to accommodate wave-plates within the early groups of the collimator. Fluid coupling in several lens groups in both the camera and collimator led to coating degradation \citep{Strydom2016}, the net result of which is poorer than anticipated blue-UV throughput and a significant increase in scattered light. The most seriously compromised elements have been replaced \citep{Crause2024}. Scattered light, if significant, has a bearing on the fiber spacing in the pseudo-slit to avoid cross-talk. We consider this below in our estimate of the fiber packing density along the pseudo-slit.

\subsection{SMI design outcomes}
\label{sec:outcome}

The SMI series (SMI-200: 0.88$\arcsec$, SMI-300: 1.33$\arcsec$, SMI-400: 1$\arcsec$.77) is designed around both the seeing distribution of the Sutherland site \citep[a median of 1.4$\arcsec$ with quartiles at 1.2$\arcsec$ and 1.8$\arcsec$;][]{Catala_2013} and flexibility to choose between spectral resolution and grasp (collecting-area--solid angle product). The first developed version, SMI-200, is optimized for higher spatial and spectral resolution at the expense of grasp. We started the fabrication process with this unit because the 200~$\upmu$m diameter fiber core of the SMI-200 is sufficiently strong to avoid frequent breakage, has sufficient shear strength to enable insertion into tightly packed apertures, yet has the lowest bending force during routing. The next highest priority development is SMI-300. This will provide the same spatial footprint and sampling as the IFU for NIRWALS, an NIR spectrograph on SALT covering 0.8-1.7~$\upmu$m \citep{nirwals}. With the upcoming red arm of RSS \citep{maxe} and the possibility of implementing a prismatic wavelength dichroic below the focal plane, this would provide IFU observations with simultaneous coverage between 0.3 and 1.7~$\upmu$m. The final planned SMI-400, based on 400 $\upmu$m diameter core fiber, provides the largest grasp, and is intended for tackling observations at low surface brightness, such as the circum-galactic medium and dwarf galaxies.

\begin{figure}[ht]
    \centering
    \includegraphics[width=0.8\linewidth]{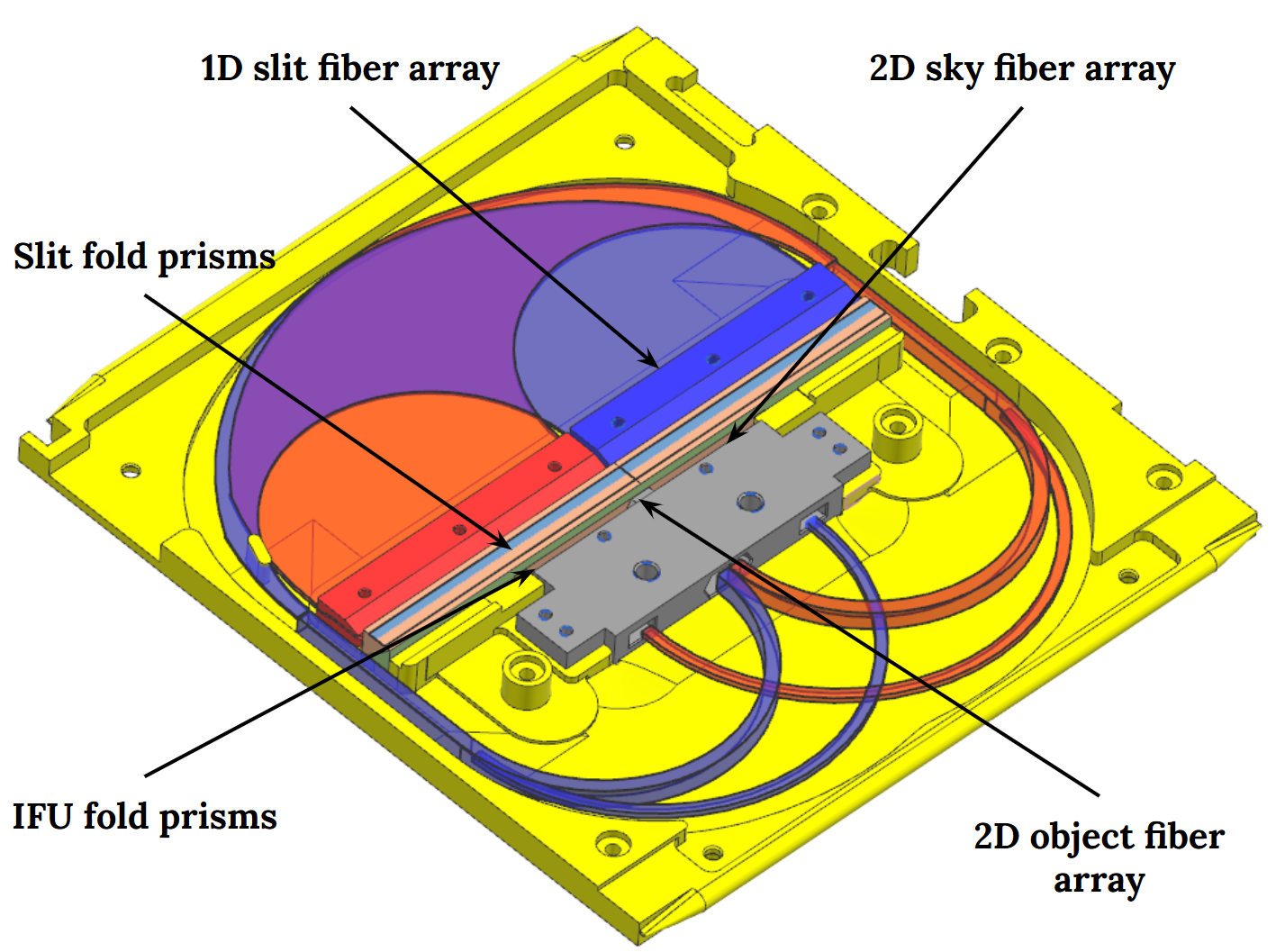}
    \caption{Mechanical layout and routing of fibers inside SMI-200 cassette, with top lid removed. Telescope light, coming up to the cassette bottom as oriented in this figure, is reflected 90 deg just before focus via a fold (`IFU') prism into the 2D object IFU and sky fiber arrays housed horizontally in a single stainless steel mount (grey block). Fibers are routed around both edges of the cassette and fed into two 1D slit blocks (stainless steel V-grooves, but color-coded red and blue), each of which is abutted to its own fold (`Slit') prism, that reflects the fiber output 90 deg back into the spectrograph collimator. The optical arrangement of prisms places the apparent fiber output focal plane parfocal with the telescope focal plane in air. For later reference, the blue slit block is referred to as the `right' block (closest to the insertion/retraction notch on the top right) with blue fibers routed around the left side of the cassette; the red slit block is referred to as the `left' block, etc. The yellow portion of the structure is machined from a single block of Uddeholm Stavax ESR soft-annealed steel. One of two guide rails for mechanism insertion can be seen at the lower right edge.}
    \label{fig:cad}
\end{figure}

\section{Optical and Opto-mechanical Design}
\label{design}

The optical design of the SMI system aims to use fibers to reformat the focal plane while keeping the focal plane location unchanged as viewed by both the telescope and spectrograph. All of this must be done within the restrictive volume of the slit-mask cassette. Given the limited space available above and below the focal plane, the incoming beam from the telescope is folded by a mirrored prism just before focus into fibers arranged into the object and sky arrays. The fibers then route 360 deg in the focal plane, where they are rearranged into the pseudo-slit; their output is redirected into the spectrograph by a second set of fold prisms.

Figure \ref{fig:cad} shows the mechanical layout of the SMI-200 assembly, with red and blue arcs indicating the fibers running from the IFU and sky arrays in the grey block (bottom right) through two side channels (top right and bottom left) to the two slit blocks (top left, color-coded red and blue for the fibers they receive). The 2D sky and object (IFU) fiber arrays, held within the grey (stainless steel) block, are butted against the IFU prism, while the 1D slit array of fibers is pushed against two slit prisms, one for each slit block (also stainless steel). The cassette base is machined from a single block of Uddeholm Stavax ESR soft-annealed steel to ensure, through a single machining exercise, that all of the IFU, V-grooves, and prism mounting surfaces are positioned within $\pm$25~$\upmu$m tolerance. Flatness of each mounting surface was achieved within $\pm$10~$\upmu$m. Single block machining was used to limit the rotational misalignment between the fiber bundles and the IFU prism to within 0.02 degree while achieving 0.01 degree between the slit assembly and the slit prisms. The machining also ensured slit-prism end-to-end effective defocus to within 40~$\upmu$m. The rigidity is provided by the cassette base of 1~mm thickness and the four corners, where the metal is 7~mm thick. Given that the available thickness is not always 8~mm at the focal plane, instead of a single lid, the entire assembly is covered by a set of three lids to mimic the volume envelope available within the letterbox mechanism. The aluminum lids are 0.3 to 1~mm thick and only enclose the fiber routing volume as well as protect the fibers from dust without providing any structural rigidity. 

\begin{figure}[ht]
    \centering
    \includegraphics[width=0.85\linewidth]{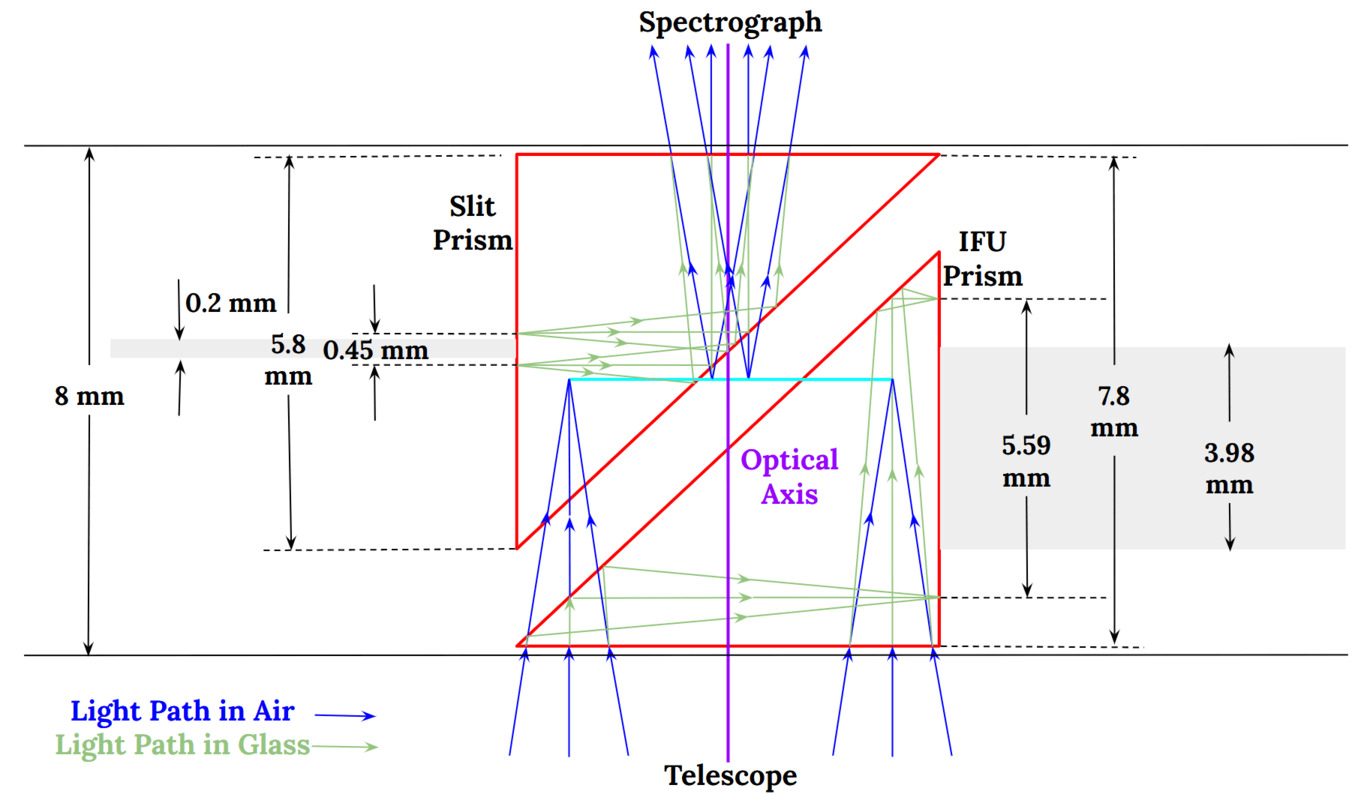}
    \caption{Coss-section schematic view of the SMI-200 central region. Telescope light from the bottom passes through the IFU prism (bottom red triangle) and is reflected via its protected Ag-coated hypotenuse to enter the 2D object fiber array sitting horizontally at focus on the right. Fibers are then routed and fed into a one-dimensional pseudo-slit (left) at the slit prism (top red triangle), which folds the light into the spectrograph collimator above. Prism catheti are coated with broadband anti-reflection coatings to reduce losses due to Fresnel reflection. For reference, the vertical dimension of the object fiber array (3.98~mm) and single fiber (0.2~mm) of SMI-200 are shown in grey at the right and left, respectively. Other critical dimensions are labeled: 8~mm outer vertical height of the SMI cassette; 7.8~mm internal height of the nested IFU and slit prisms; 5.8~mm prism cathetus; and 5.59~mm prism clear aperture.}
    \label{fig:opp}
\end{figure}

Figure~\ref{fig:opp} illustrates the light path (blue for the path in air and green for the path folded by the IFU and Slit prisms) through the cross-section of the fold prisms (red triangles). All prism catheti are AR-coated, yielding $<1$\% loss per surface over RSS operable wavelength range given the f/4.2 beam from the corrected SALT primary. Hence, the fibers and prisms are air-gapped. We considered optically coupling the fibers directly to uncoated fold prisms, but this introduced either alignment challenges for the application of index-matched epoxy or longevity risks for the application of index-matched gel since there was insufficient volume to provide adequate encapsulation to protect against degradation in the harsh telescope environment. In fact, because the vendor could not provide AR coatings on only one prism leg, optical coupling would have yielded little gain. While \cite{drory} have shown that it is possible to AR-coat assembled fiber bundles and slits, the inward facing orientation of the SMI IFU and slit assembly and the fragility of the folded assembly used for polishing (\S~\ref{sec:slit}) precluded this as a safe option for the SMI series.

In the absence of the bottom fold prism, the telescope beam would focus on the cyan line (representing the telescope focal plane in air). Instead, the rays follow the green path in the N-BK7 prism glass, reflected off the protected Ag-coated hypotenuse, exiting through the other AR-coated surface. For reference, protected Ag has reflectivity of (90,87,88,94,97.5,99.5)\% at (350,370,400,500,700,900)~nm. The separation between the prisms and their dimensions are designed so that the projection of the IFU and Slit prism ray paths coincide with the telescope focal plane in air. This `parfocal' design eliminates the requirement for adjustment to the telescope and spectrograph collimator foci during operation, which would be difficult to achieve without the introduction of additional focus procedures.

The available thickness of the SMI cassette along the optical axis is found to be mostly 9 mm. Although in a couple of areas this thickness limit is down to 7 mm, the thickness available around the slit is 9 mm. The thinner areas correspond to features that accommodate (a) a slit mask ID reading barcode sensor, and (b) a magnetic strip for attaching the cassette to the elevator mechanism. To ensure smooth operation around these features, we chose 8 mm thickness of the cassette. As discussed earlier, the cassette base provided the rigidity while the thinner lids act as dust covers. The volume around the slit is used to fold the incoming optical beam using the prisms, for the IFU and the slit. The prism dimension is defined by the need for a parfocal (in air) optical system, implementation of a safe mechanism, and available volume as described earlier. The SMI and longslits are made parfocal to minimize refocusing during operation between longslit and SMI exposures. Use of longslit is necessary for SMI observation. Due to the space constraint, SMIs do not have any imaging surface for target acquisition. The targets are expected to be acquired on a reflective surface of a longslit mask by imaging the surface on a slit view camera. Then the longslit mask is replaced by the SMI cassette, which would then position the target on a specific fiber within some mechanical accuracy calibrated during commissioning. In addition, the SMI cassettes are expected to go through several storage, travel, and insertion, each of these activity would pose a potential risk to the optical surface of the prisms. Thus, we designed the entry and exit surface distance of the two sets of prisms to be 7.8 mm, as shown in figure \ref{fig:opp}. This choice of dimension enabled us to provide a 0.1 mm gap between the cassette top and the prism exit surface and the prism entry to the cassette bottom surface.

The dimension of the prism cathetus is defined based on the material, and to conserve the parfocal plane viewed by the telescope and spectrograph. The choice of material for the prisms was decided to be fused silica, based on availability, shortest lead time, and the ability of the manufacturer. This, coupled with the requirement of a parfocal system, fixed the prism catheus dimension to 5.8 mm with a separation of 2 mm along the optical axis. Of the 5.8 mm, the clear aperture was defined as the central 5.59 mm to ensure minimal vignetting on the marginal ray entering the IFU prism. A larger clear aperture would require either increasing the prism cathetus dimension or moving the prism laterally, both of which would disable the parfocal feature. The slit prism could have been made smaller, but kept similar dimensions to those of the IFU prism to minimize non-recurring fabrication cost.

Apart from the requirement of clear aperture and prism cathetus dimensions, our tolerancing analysis showed prism angles must be within $\pm$0.1 degree of the expected values. We also requested a protected silver coating for $>$95\% reflectivity on the hypotenuse and $<$1\% loss on the AR-coated cathetus surface. Finally, given our ability to terminate the fiber blocks with high surface finish, we defined the required surface roughness to be $\uplambda$/4 with 10 $\upmu$m flatness. We kept a 1 mm space on each side of the prisms for machine handling during the fabrication. \footnote{Rocky Mountain Instruments - https://www.rmico.com/}

The on-sky equivalent 24.6~$\arcsec$ footprint defines the limiting height of the IFU bundle, with the only limit in the other dimension being the slit-length itself. Ultimately, the IFU footprint area is limited by the total number of fibers the slit assembly can accommodate, but the height of the footprint is dictated by the clear aperture of the AR coatings on the prism catheti.

The following design considerations were used to determine the detailed complement of the SMI-200 fibers, and how they were distributed between object and sky arrays and within the object array: Cross-talk constraints in the fiber pseudo-slit determine the total fiber count (\S~\ref{sec:crosstalk}), while minimizing random errors from sky subtraction set a minimum allocation of sky fibers (\S~\ref{sec:skyran}). Compelling IFU geometries (\S~\ref{sec:array}) and sampling SALT's non-telecentric focal plane (\S~\ref{sec:NT}) determine the final distributions between object and sky (\S~\ref{sec:skytel}) as well as the detailed mapping between IFU and slit.

\subsection{Cross-talk constraints on fiber spacing in the pseudo-slit: total fiber count}
\label{sec:crosstalk}

The total number of fibers in SMI-200's pseudo-slit is 336. This was determined via simulations by the \textit{ab initio} requirement that there be a 1:10 trough-to-peak contrast in the fiber trace at all detector field points to minimize cross-talk between fibers adjacent in the pseudo-slit. (Such contrast also minimizes the burden in automatic trace identification algorithms, but this is of secondary importance.) This requirement was based on the sensible argument presented for the MaNGA system \citep{drory} that the cross talk in the spectrograph detector plane be less than that introduced by seeing in the telescope image plane. For MaNGA, the average cross-talk levels across all field points in the detector focal plane was roughly 1:10, with corrected values better than 1:20. Our simulutions do not include as-built optical alignment errors or scattered light and consequently, as we show in Paper II, we do not achieve 1:10 contrast on sky. However, with the relatively smaller SMI-200 fiber foot-print on sky and poorer seeing at SALT compared to MaNGA, we are still able to keep cross-talk between fibers in the spectrograph focal-plane sub-dominant with what we will refer to as our idealized 1:10 contrast requirement.

To arrive at the fiber spacing requirements, we used Zemax to compute the encircled energy (EE) distributions at the RSS detector focal-plane for a wide range of grating configurations and a two-dimensional grid of detector field points for each configuration. The input image was the 200$\upmu$m fiber core, assuming a uniform input surface-brightness. Given that aberrations are modest, the output surface-brightness profile is close to a top-hat function with rounded edges and wings, the latter being the critical ingredient for estimating cross-talk. These EE distributions were used to construct the fiber trace profile in the pseudo-slit dimension (integrating the monochromatic distributions along wavelength). These profiles were interpolated in both field positions on the detector and then pixelated (summed over the 15$\upmu$m detector pixels) to simulate the fully populated pseudo-slit as would be observed, e.g., in a continuum-lamp exposure. The simulations allowed for the fiber spacing along the slit to vary as needed to modulate the trace contrast. Since the RSS optical performance is symmetric around the optical axis, we simulated only half of the pseudo-slit. 

From a series of simulations for different spectral locations and spectrograph configurations, we found optimal fiber separation in four different zones, each $1\arcmin$ (or roughly 27~mm) wide, extending from the center (Zone 4) to the edge of the slit (Zone 1). The choice of 4 discrete zones came about because we were considering up to 8 V-groove blocks, as had been done in NIRWALS \cite{nirwals}. In the end, we adopted 2 V-groove blocks, which decreases sampling of scattered light in the inter-block regions but does enable us to increase the total fiber count. 

\begin{figure}[ht]
    \centering
    \includegraphics[width=\linewidth]{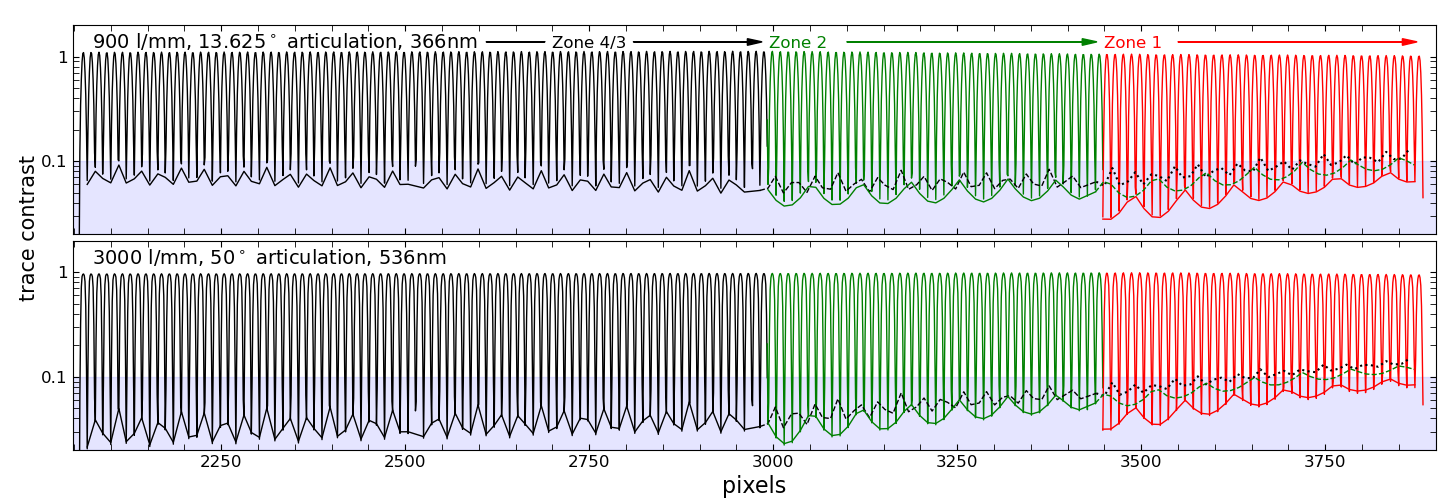}
    \caption{Fiber trace simulations of the as-built SMI-200 (top, regular periodic curves) and trace contrast (bottom curves) for half of the fiber pseudo-slit and for wavelength extrema in two spectrograph configurations (as labeled in top and bottom panels). Trace contrast, defined as the trough (fiber gap) to peak (fiber center) ratio, hugs the bottom of the trace when the peaks are scaled to unity, as done here. Fiber center spacing varies from 300~$\upmu$m in Zones 4 and 3 to 315~$\upmu$m in Zone 1, as described in the text to keep contrast $<0.1$. The degraded contrast of the higher density packing of the inner slit regions, had this density been extended to larger field angles (larger pixel values), is shown by suitably colored dashed and dotted lines in Zones 2 and 1 (black for Zone 4/3 spacing extended into Zones 2 and 1; green for Zone 2 spacing extended into Zone 1). These two spectrograph configurations are representative of the most constraining performance of RSS image quality. With this fiber spacing, most grating configurations over most of their covered wavelength range have simulated contrast $<0.01$.}
    \label{fig:fullcontrast}
\end{figure}

Because RSS image quality is dependent on both wavelength and grating articulation angle, we simulated a wide range of spectrograph configurations, including the VPH900 at 13.625$^\circ$, 17$^\circ$ and 19.95$^\circ$ with broad wavelength coverage between 366 and 900~nm; VPH1800 at 39$^\circ$ and VPH2300 at 50$^\circ$, spanning the H$\alpha$ region from 634-755~nm and 623-699~nm, respectively; as well as the VPH3000 at 50$^\circ$ covering 478-536~nm. RSS image quality degrades rapidly redward of 850~nm, so we do not consider this wavelength regime as a constraint on the fiber slit spacing. Outside of this wavelength regime, the most stringent constraints come from the edge of the field (end of the pseudo-slit) for the 50$^\circ$ configurations at the red wavelength limit (536~nm for VPH3000 and 699~nm for VPH2300), and from VPH900 at 19.95$^\circ$ at the blue wavelength limit of 605~nm. However, the worst contrast over the full slit length is seen in the VPH900 settings for 13.625$^\circ$ at the wavelength extrema (366~nm and 674~nm). Examples of two limiting cases are shown in Figure~\ref{fig:fullcontrast}. By varying the fiber spacing in each of the 4 zones, we were able to keep the typical contrast below 1:10. The two central zones (Zone 4 and 3) host 43 fibers each, while the outer zones (Zone 2 and 1) contain 42 and 40 fibers each. This translates to 300, 306, and 315 $\upmu$m fiber spacing, center-to-center, in Zone 4/3, Zone 2, and Zone 1, respectively. Thus, each half of the slit hosts a total of 168 fibers. The full fiber pseudo-slit uses 102.5~mm of the available 105.6~mm (8$\arcmin$) slit length, with the difference due to a $\sim$1~mm (gap between adjacent V-grooves between the two blocks (at the slit center), and $\sim$1~mm gaps at each slit end due to the $\sim5^\circ$ NT tilt of the outermost fibers across the 11.2 mm block width. Referring to \cite{rss_ops} for a general layout of the RSS detector, the SMI-200 slit maps onto roughly 3725 of the 4096 pixels in the slit dimension, leaving roughly 130 and 240 pixels on either side of the slit nominally unilluminated. The inter-block gap is roughly 50 pixels. In Paper II we demonstrate how the existing gaps between fibers in the pseudo-slit can be used to make a correction for this scattered light.

\subsection{Minimal sky fiber allocation}
\label{sec:skyran}

As provided by equation 4 in \cite{Bershady_2004}, based on a signal-to-noise merit function, the optimum number of sky fibers for a total fiber complement of 336 is 12 sky fibers in the detector-limited regime and 17 in the background-limited regime. Except for short exposures at the highest grating angles, SMI-200 observations should predominantly fall in the latter regime. We take 17 fibers (5\%) to be the \textit{minimum} sky complement for SMI-200 as a boundary condition in our design.

\subsection{Object IFU array size}
\label{sec:array}

We have chosen an extended hexagonal shape with hexagonal packing of fibers within the IFU for the SMI series to accommodate cases for the larger fibers where a regular hexagonal aperture exceeds the vertical clear aperture allowed by the fold prisms. While the clear aperture constraint is not a limitation for the smaller SMI-200 fibers, the total fiber count and sky fiber allocation do preclude a fully packed, regular hexagon that exceeds 9 rings, while a regular hexagon of 9 rings would use only 271 fibers, leaving a sub-optimally large number (19\%) for sky. 

\begin{figure}[ht]
    \centering
    \includegraphics[width=\linewidth]{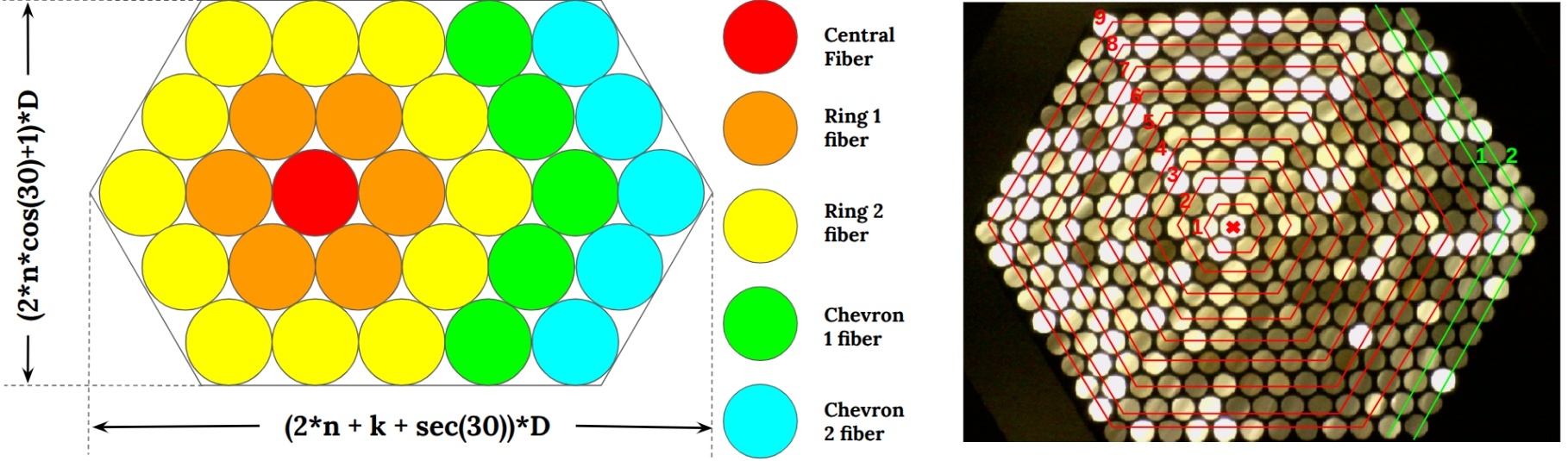}
    \caption{Definition of rings and chevrons in an extended hexagonal packing (left). In this example, there are n=2 rings and k=2 chevrons. The SMI-200 (right) has n=9 rings and k=2 chevrons, and a buffer OD of D=240~$\upmu$m.}
    \label{fig:ringrow}
\end{figure}

The total number of fibers in a generalized (extended) hexagon is $\rm [3 \times n \times (n + 1) + 1] + k \times (2 \times n+1)$ for n rings and k additional chevrons. Ring and chevron definitions are described in Figure~\ref{fig:ringrow}. The dimension of a hexagonal aperture is defined by the number of rings and rows. The longest corner-to-corner width is $\rm [2 \times n + k + \sec(\pi/6)] \times D$ where D is the outer diameter (OD) of the fiber buffer. The smaller side-to-side dimension is $\rm[2 \times n \times \cos(\pi/6)+1] \times D$. The addition of the chevrons allows for an increase in the IFU area and object fiber count that provides a more optimal distribution of object and sky fibers. 

For SMI-200, we chose n=9 (271 fibers) and k=2 (38 fibers) for an allocation of 309 object IFU fibers, leaving 27 fibers (8\%) for sampling sky. This yields an elongated hexagon of 19 $\times$ 21 fibers with physical dimensions of 3.98 $\times$ 5.08 mm including clad and buffer, corresponding to $\rm b/a=0.783$ of a circular disk (e.g., a spiral galaxy) observed at an inclination of $\sim$38~deg. The on-sky IFU footprint is $17.6\arcsec\times22.5\arcsec$ as shown in Figure~\ref{fig:smi}.

\subsection{SALT non-telecentricity: mechanical solutions}
\label{sec:NT}

\begin{figure}[ht]
    \centering
    \includegraphics[width=0.8\linewidth]{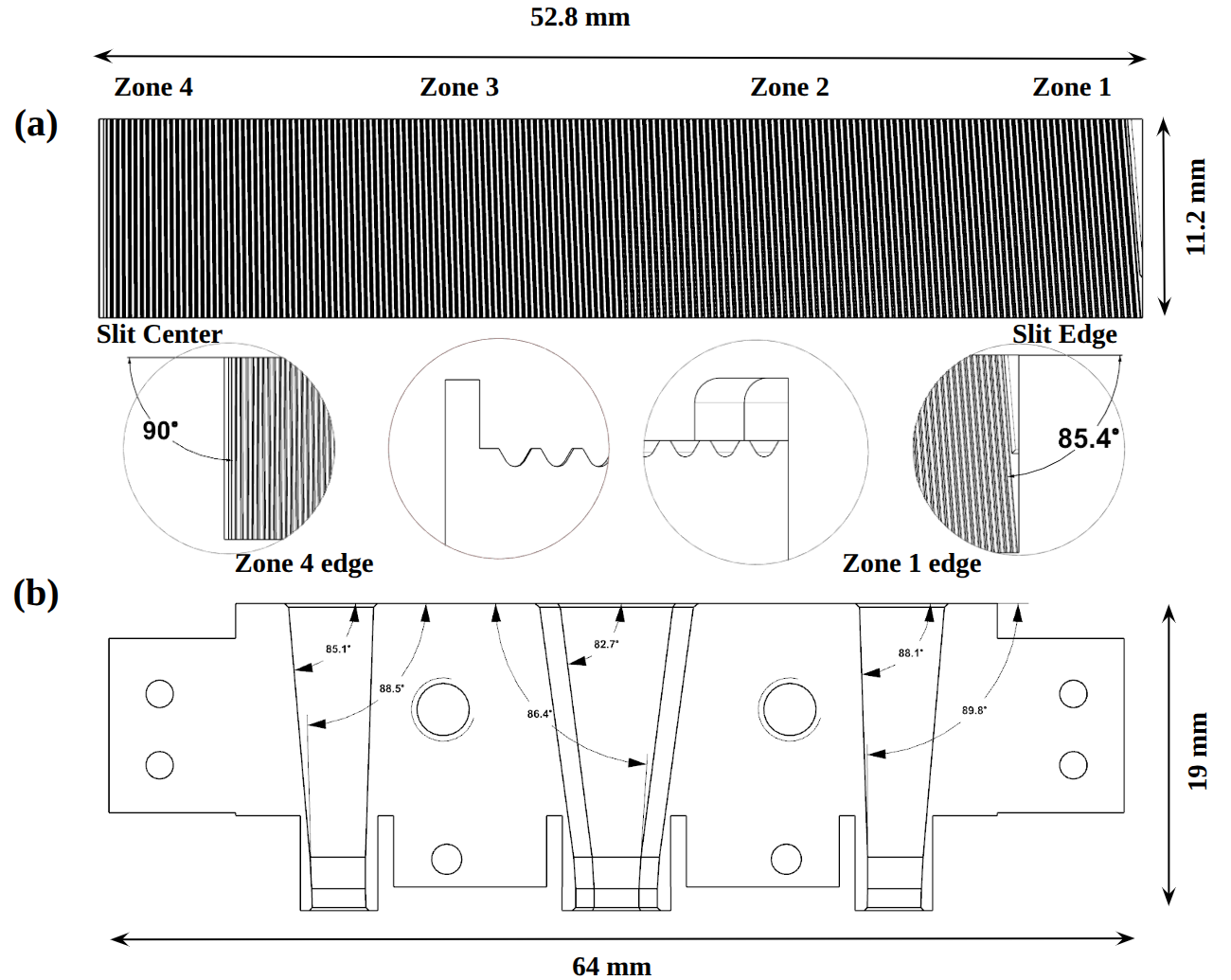}
    \caption{Optomechanical design of fiber array holders of SMI-200. The top panel (\textbf{a}) shows half of the linear array holder (one of two V-groove blocks) that defines the fiber pseudo-slit. An expanded, perpendicular view of the slit edge and center shows how V-grooves are fanned to position fiber output beams at increasing \textit{out}ward angle with field position to mimic the non-telecentric telescope focal plane shared by the RSS collimator. Bottom panel (\textbf{b}) shows a cross-sectional view of the bi-dimensional fiber array holder, the central aperture holds the object sampling fibers, while the side apertures hold sky sampling fibers. The sky apertures are slightly tilted {\it in}ward, again with the intent to match the non-telecentric input of the telescope.}
    \label{fig:optomech}
\end{figure}

Several practical consequences flow from SALT's non-telecentric focal plane. The first and perhaps most obvious is that the fiber pseudo-slit must have the fibers splayed at an increasing angle with respect to the slit normal at increasing distance from the field center; RSS is designed to accept the NT of the focal plane. This is achieved mechanically through the design and fabrication of tilted V-groove blocks, illustrated in Figure~\ref{fig:optomech}, panel (a).

Given that the IFU is centered on the SALT focal plane, the NT distribution for IFU fibers is modest, but non-negligible. Due to the fiber azimuthal scrambling, NT injection of the optical beam into the fibers leads to additional geometric focal ratio degradation (gFRD). We have no mechanical solution for this issue, but we do ameliorate the impact via a carefully constructed mapping protocol (below) that impacts our total sky fiber allocation. 

More significant is the impact on the sky fibers since we desire to place them at some distance from the object IFU to well-sample the sky about extended sources. SALT's SAC does introduce significant field-dependent vignetting at prime focus, but this is minimal within the inner 90$\arcsec$ radius in field angle. For this reason, we limit the sky fibers to be placed in two groups, each at a distance of $90\arcsec$ from the center of the object bundle along the slit direction. Since the sky fibers would have large NT angles ($90\arcsec \times 76 \ \text{times}$ the radial distance), their vignetting profile would be different from the object fibers. This difference would render the sky subtraction suboptimal. Thus, we modified the sky fiber NT angle distribution to mimic the object fibers by introducing a counter-angle to the sky fibers opposite to that of the NT angle, as shown in Figure \ref{fig:optomech} panel (b).  Unfortunately, the fabricated versions of the V-groove slit blocks had the non-telecentric fan angles reversed in sign from the intended values. This leads to significant vignetting as detailed in Paper~II; this vignetting occurs in the spectrograph due to offsetting of the far field footprints of the fibers relative to the instrument pupil.

\subsection{SALT non-telecentricity: sampling and mapping solutions}
\label{sec:skytel}

\begin{figure}[ht]
    \centering
    \includegraphics[width=0.9\linewidth]{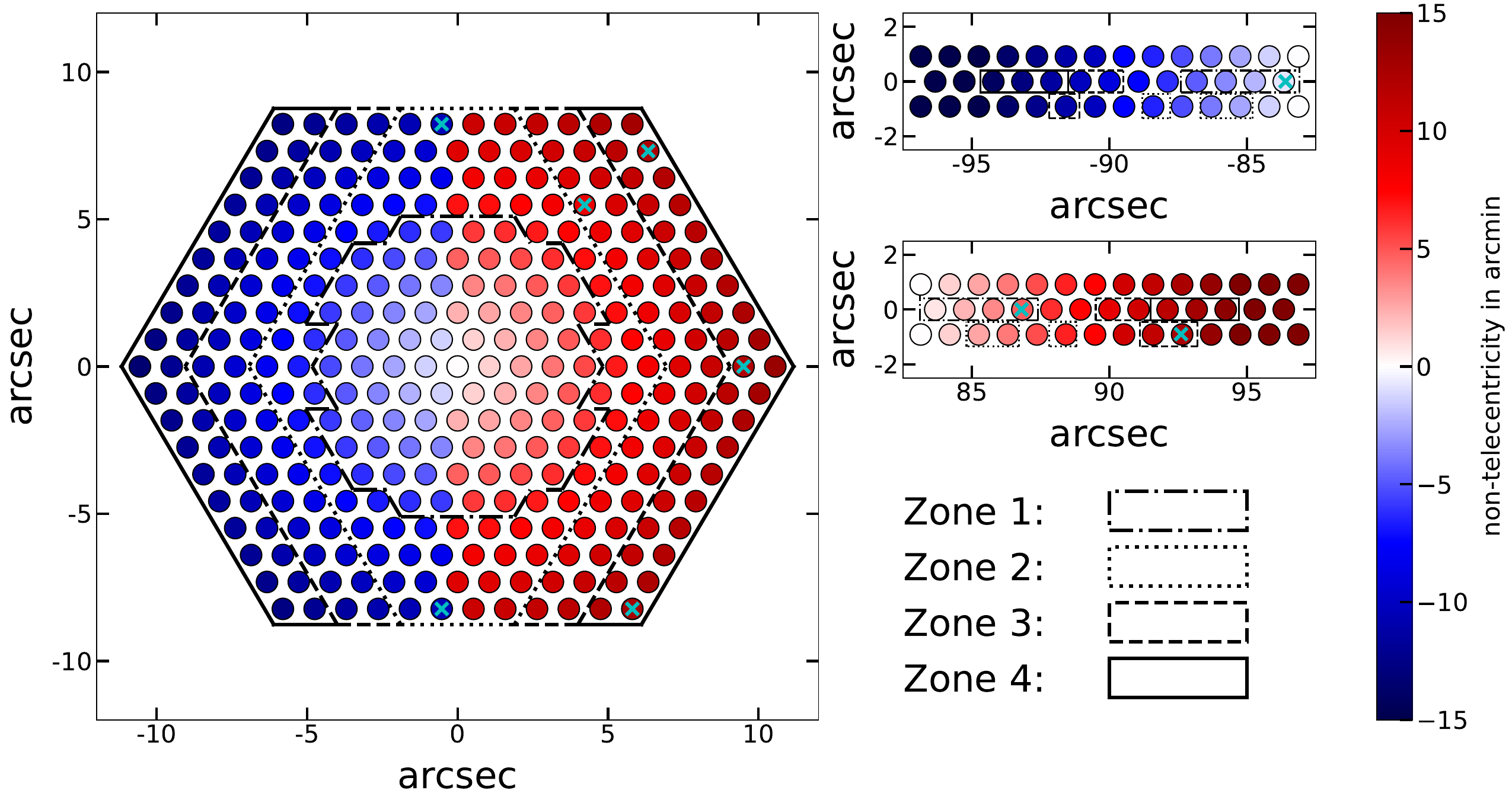}
    \caption{Map of fiber non-telecentricity on sky in the object IFU (left panel) and sky arrays (right panels). Blue and red define the non-telecentric angle direction in the right and left slit respectively (reference Figure~\ref{fig:cad}). Four different zones (1 to 4) within the IFU are chosen from the center to the edge, and are mapped to the corresponding zones in the slit, where zone 4 is in the slit center, and zone 1 is at the slit edge. Sky array fibers with matching non-telecentric angles are mapped to the corresponding slit zones. Fibers marked with `$\times$' were not usable (see \S~\ref{sec:abstp}).}
    \label{fig:ntdist}
\end{figure}

The distribution of IFU fiber NT is mapped in Figure~\ref{fig:ntdist}. We split the fibers into non-telecentric zones (rings) and pair them with sky fibers selected to have similar NT angles. These object and sky fibers are grouped together and fed to contiguous linear zones of the pseudoslit to ensure that fibers with similar NT experience a similar optical path inside the spectrograph and, hence, similar vignetting, grating efficiency, and optical throughput.  We bin fibers into NT zones of $<$0.1 deg so that beam area increases (inversely proportional roughly to the throughput decreases from vignetting) due to gFRD for the f/4.2 beam is $<3$\%. Except for the innermost zone, the non-telecentric angular range is closer to 0.03 deg, or a 1\% increase in beam area. This yields 4 zones. 

Given the IFU symmetry and the need to minimize bare fiber overlap in the routing, the 4 zones are mapped to both halves of the slit, for a total of 8 zones. We aim to sample each of these eight zones with three sky fibers spaced uniformly along the zones in the slit to map changes in the spectrograph LSF with field angle at nearly constant beam profile (spectrograph illumination). This yields a total of 24 required sky fibers rather than the 17 predicted from only optimizing SNR (\S~\ref{sec:skyran}). For our IFU geometry and total fiber complement, we had initially allocated 27 sky fibers, very close to this number.

The SALT focal plane has negligible changes in vignetting over the IFU field of view. Moreover, the spectrograph optics also do not produce vignetting over the full 4$\arcmin$ entrance slit for an appropriately non-telecentric, f/4.2 input beam. However, fiber FRD will introduce vignetting within the spectrograph. This vignetting depends on both the amplitude of the FRD (e.g., the fraction of light contained within an f/4.2 beam) as well as the field position within the entrance slit, with more vignetting for larger amounts of FRD (faster input beams at a given field position) and for slit positions farther from the field center (at a given amount of FRD). Since the outer regions of galaxies have lower surface brightness, and the outer fibers in the SMI-200 IFU suffer more from gFRD (due to the larger NT angles), we compensated these effects by mapping the IFU outer fibers into the inner regions of the slit (zone 4), and vice versa, as indicated in Figure~\ref{fig:ntdist}. In this way, we aimed to minimize the dynamic range in SNR for fibers within the entire assembly. As it turns out,  fibers mapped closer to the slit center undergo progressively tighter bend radii during their routing on the slit-block portion of the SMI cassette. This can be understood by inspecting Figure~\ref{fig:cad}. These tighter radii induce somewhat more FRD, mostly offsetting the throughput gains obtained by their more central field-positions. We measure and model these effects as manifested in our mapping choice in \S~\ref{sec:performance}.

\section{Development Methodology}
\label{sec:assembly}

\subsection{Aperture Selection}
\label{sec:apsel}

The IFU geometry was achieved by constructing a tapered hexagonal aperture, akin to \cite{drory} for the MaNGA IFUs, into which the fibers were inserted to form a naturally dense array. For ease of insertion, the aperture has multiple chamfers with decreasing angle to form the taper, leading to a straight section of 3 mm in length for glue bonding, as shown in Figure \ref{fig:optomech} (panel b). Molex/Polymicro\footnote{\url{https://www.molex.com/en-us/part-list/optical-fibers}} made broadband fiber FBP200220240 (part no. 1068022592) fibers with a numerical aperture of 0.22 are used with core:clad:buffer having a diameter ratio of 200:220:239 in $\upmu$m. The tolerance on the buffer diameter is defined by the manufacturer to be $\pm$5~$\upmu$m but in practice it is much smaller within a single draw. Instead of measuring the as-delivered fiber diameter, we chose to empirically find the clearance required above the nominal aperture size to ensure ease of insertion yet tight packing. We chose this empirical route because fiber diameter uncertainties are only one factor determining the optimal aperture; surface roughness was also a factor. For uniform and regular packing of fibers, the dimensions of the aperture need to be precise, yet with sufficient clearance to allow the fibers to be inserted without damage. We estimated the clearance on the corner-to-corner direction to be $\upsigma_{\rm major} = \upepsilon\times\sqrt{\rm2\times n + 1 + k}$, the clearance on the shorter side to be $\upsigma_{\rm minor} = \upepsilon\times\sqrt{\rm2\times n + 1}$, where $\upepsilon$ is the uncertainty (or variation) in the fiber diameter.
 
The IFU and sky apertures were machined via a plunge EDM process. To find the aperture dimension with the proper clearance for the IFU, we fabricated test apertures with different integer multiples of $\pm\upsigma$ to the nominal dimensions. For the purpose of setting these aperture steps we chose $\upepsilon=5~\mu$m even though this is larger than what is expected. The test apertures were measured under a shadowgraph to ensure the specified dimensions. We packed 309 short fiber lengths into each test aperture, and for those where the insertion was possible, terminated them with glue, and polished them sufficiently to clearly resolve all individual fiber locations. The test apertures were ranked based on ease of insertion and the tightness and regularity of packing. The best aperture, used for the final article, was found to have the dimensions of 5.12 mm$\times$4.02 mm, larger than the nominal size by 45 $\upmu$m in both dimensions or roughly $2\upsigma$.\footnote{We do not believe there is inference from this result on the true fiber OD variation; rather, the aperture over-sizing that optimized the trade-off between ease of fiber insertion and packing regularity was likely also dependent on surface-friction due to aperture surface roughness and area.}

\begin{figure}[ht]
    \centering
    \includegraphics[width=0.85\linewidth]{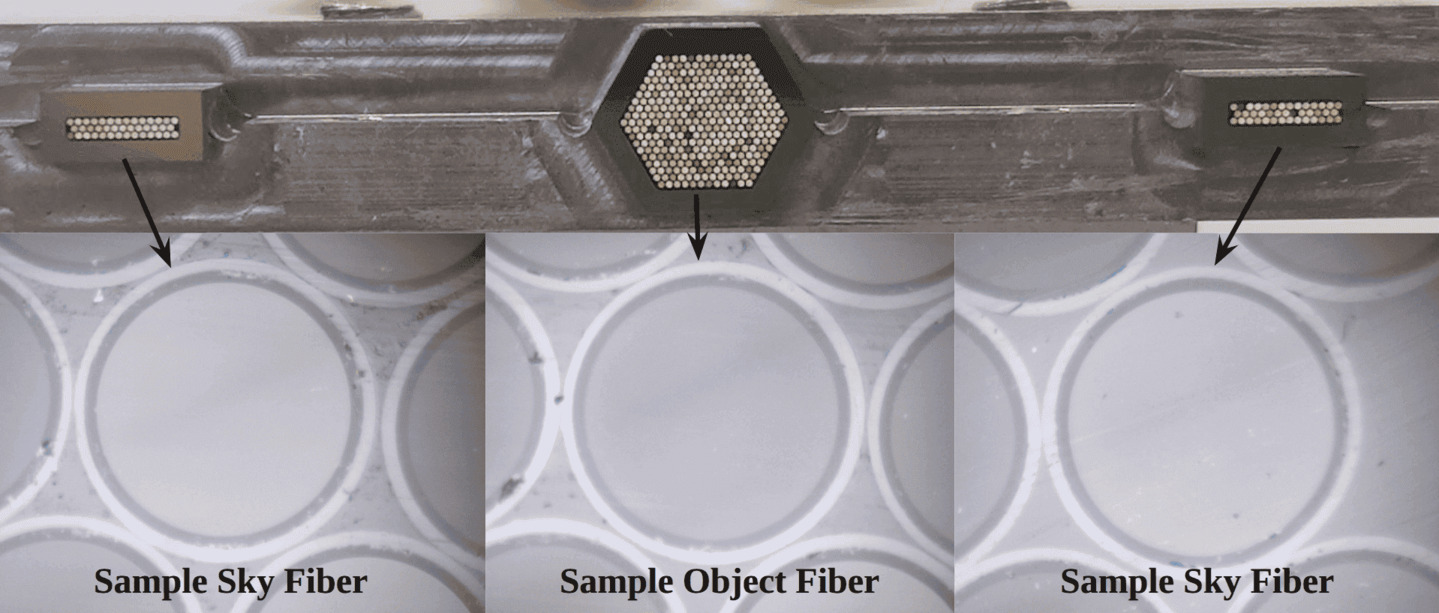}
    \caption{IFU and sky array block after polishing (top). Intensity variations seen here are due to the vagaries of the fiber illumination in the lab. Representative sample regions of fiber surface after polishing (bottom panels) show the excellent final surface finish. This figure is a modified version of figure 2 in \cite{chattopadhyay2024}.}
    \label{fig:polish}
\end{figure}

Sky apertures are rectangular in shape to minimize the total fiber count while providing sufficient field length to sample the same range of NT as the object IFU. Rectangles were formed to accept 41 fibers in three rows of 14, 13, and 14 hexagonally packed fibers, as shown in figure \ref{fig:polish}. The nominal dimension is 3.36 mm$\times$0.65 mm, which we increased by $2\upsigma$ to  3.38 mm$\times$0.66~mm in final fabrication.

\subsection{IFU and sky array fiber bonding}
\label{sec:bonding}

The object and sky fibers were bonded in their respective apertures using Epotek 301\footnote{\url{https://meridianadhesives.com/products/epo-tek-301/}}, a two-part room-temperature curing epoxy with low viscosity. The epoxy was pre-cured for one hour to achieve the required viscosity to ensure gaps between fibers are filled, but the capillary action was not too strong to wick the glue beyond the 3~mm aperture entry length described above. During the pre-curing process, air bubbles were eliminated (to ensure consistent filling of fiber gaps) by allowing them to naturally percolate out of the epoxy held within a syringe. The pre-cured epoxy was applied to the fiber tips pushed 1~cm beyond the aperture over a width of 5~mm using a spare fiber to distribute and work the glue between all bundle fibers. The adhesive-filled fibers were then pulled back into the aperture, leaving 2~mm glued extent beyond the aperture end. This assembly was cured for 24 hours.

\subsection{IFU and sky array polishing}
\label{sec:skypol}

The fibers were polished by mounting the fiber assembly onto a custom polishing arm \citep{polisher} designed and fabricated for an off-the-shelf Ultra-tech Ultrapol Basic lapping machine\footnote{\url{https://www.ultratecusa.com/product/ultrapol-basic/}.}. The latter has a 200~mm diameter circular polishing platen in a horizontal orientation that rotates at a selectable rate that can be varied continuously up to 10,000 rpm. The custom arm enables careful control of load application, ensures surface alignment to within 10 $\upmu$m, and provides the ability to inspect the polished surface during the polishing process without disturbing surface alignment. The polishing station has a three axis stage to define tip-tilt on the fiber surface. A gauge block was used to remove any wedge by adjusting the two reference planes of the V-grooves with respect to the polishing platten at different polishing locations, both radially and azimuthally. We polished the surfaces of all three apertures simultaneously using 200~mm circular diameter polishing discs with 15, 5, 3, and 1 $\upmu$m grit\footnote{3M$^{\rm TM}$ product packaged by Ted Pella Inc., \url{https://www.tedpella.com/}.} for 5, 20, 60, and 120 min, respectively. The 15 and 5 $\upmu$m grits were from silicon carbide to perform quicker polishing, while the finer grits were based of aluminum oxide, which provides slower but more controlled polishing. The polishing parameters were found to be very similar to those described in \citep{polisher}. The load was adjusted to 100~gm for all grits except 1 $\upmu$m grit disks, where the load was reduced to 50~gm. After each grit, the object and sky arrays were inspected for scratches and digs larger than the grit size. If no such features were found, polishing proceeded to disks using smaller grit. Figure \ref{fig:polish} shows the resultant fiber surface after polishing with 1 $\upmu$m grit. All 336 fibers were inspected to ensure the number of digs and scratches is less than five over the entire array.

\subsection{Astrometric specification of the IFU array}
\label{sec:astrom}

We measured the positioning of the fibers with an automated script \citep{gaedie2025} and found the distribution width of random fiber positioning error to be $\pm 2.7 \ \upmu$m. The script measured the positions of each fiber by comparing against an ideal grid with fiber-to-fiber separation of 0.239 mm. Individual fiber images were extracted based on the reference grid, and the fiber centroid was measured. The centroid coordinate was compared to the grid coordinate to measure the error. The grid origin was then moved around to find the smallest error distribution width. A typical gap between the fibers after polishing, as shown in Figure \ref{fig:polish}, is on the order of $5–7 \ \upmu\text{m}$, which contributes to the random positioning error. This error is consistent with the slightly larger ($45 \ \upmu$m) dimension used in the aperture sizing. The positioning error of any fiber is defined to be within $0.1\arcsec$ on a reference grid centered on the central fiber of the object array.

\subsection{Fiber routing between fiber head and slit}
\label{sec:routing}

Once terminated, object and sky fibers were grouped into NT zones, as discussed in \S~\ref{sec:skytel}, and fed to the slit assembly while covered in a PVC jacket for protection. The PVC jackets are $\sim$100~mm long with a wall thickness of 0.3~mm. Each group of fibers was routed through its own PVC jacket to help maintain the order of fibers on the slit. Sky fibers from the same zones were added to form the fiber groups. This routing and the positioning of the fibers were by far the most difficult part of the assembly, given their fragility, yet the substantial collective force from the tight bend diameters. A further complication was the need to assemble the fiber into the pseudo-slit using V-groove blocks oriented and positioned in their final location. This location requirement ensured proper fiber lengths and paths. A final complication was the polishing of the fiber slit. Noting the tight, facing configuration of the slit and IFU about the fold prisms, it is clear that polishing the slit in the fully assembled system with tightly routed fiber is non-trivial. This polishing required a carefully executed process.

\begin{figure}[ht]
    \centering
    \includegraphics[width=0.85\linewidth]{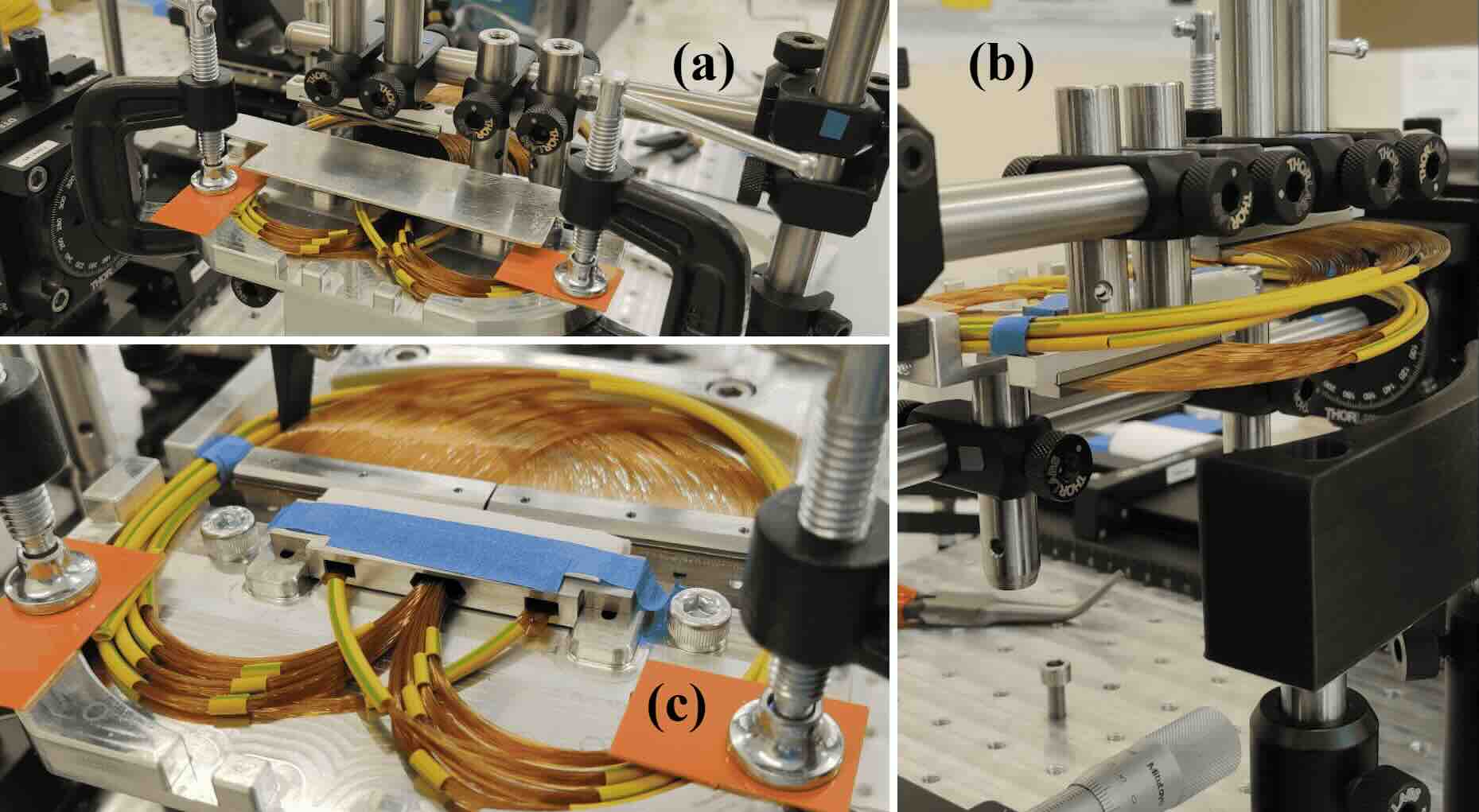}
    \caption{Routing of fibers from the IFU end to the slit end. Top-left panel (\textbf{a}) and right panel (\textbf{b}) show the relative positioning of V-grooves during fiber insertion into the `right' block, defined in Figure~\ref{fig:cad}. Bottom-left panel (\textbf{c}) shows the fibers routing after both V-grooves have been positioned after assembly but before polishing. The left routed fibers entering the right V-groove block have a tighter bend, leading to higher FRD and vignetting.}
    \label{fig:routing}
\end{figure}

The solution we devised was to design and fabricate an assembly jig with the same internal dimensions and bending arcs as the cassette (Figure~\ref{fig:cad}), but split into two pieces with a 10 mm gap perpendicular to the slit dimension between mounting blocks for the IFU and V-groove blocks. These two half-cassettes each had ample fixture to be held in position with posts on an optical bench with translation and rotation stages, such that they could be actuated relative to each other in height, separation, and angle. These different relative positions were used in the polishing phase, as described below. For the routing and slit-assembly phase, the two half-cassettes were held fixed so that they mimicked the final cassette internal dimensions and relative orientation of IFU and sky array head and slit blocks (Figure~\ref{fig:routing}). With the IFU and sky array head attached to its half-cassette mount, the fibers were then routed to the V-groove blocks forming the pseudo slit in the other half-cassette mount.

To minimize fiber bend diameters, half the object and sky fibers were routed around each side of the cassette, as seen in Figure~\ref{fig:cad}, and then positioned into the slit-block on the opposite (far) side of the cassette. From trial (and much) error, we found that splitting the IFU fibers into their routing paths by their horizontal position within the IFU was less prone to inducing fiber damage because this prevented tightly-packed fibers from crossing over each other where they were likely to rub.

During the insertion of the fibers into the V-groove blocks forming the pseudo-slit, there was significant spring force from each fiber due to its bend. Using the V-groove blocks' lids, the gap between the grooves and the lid was accurately spaced with shims to ensure smooth insertion of individual fibers while not allowing them to jump into a neighboring V-groove. During fiber insertion, each V-groove assembly (left and right) was held at what will be its final location in the SMI-200 cassette (relative to the IFU), one assembly at a time. This ensures that the fiber lengths for each route were accurately set. The insertion process started with placing Zone 4 fibers (at the outskirts of the IFU) into the center of the pseudo-slit block, and ended with Zone 1 fibers (in the IFU center; refer to Figure~\ref{fig:ntdist}). Since the fibers were inserted into the slit block one at a time, this ordering was chosen simply to have open access to the remaining open block.

Once the routing and insertion of a half-slit assembly were complete, the fibers were bonded and cured. After curing, any excess fiber length beyond a few mm from the slit end was trimmed for polishing preparation. The fully-terminated half-slit assembly was then lifted up in a controlled manner to provide vertical clearance to insert fibers in the other half-slit assembly, as shown in panels (a) and (b) of Figure \ref{fig:routing} for the left V-groove block (as defined in Figure~\ref{fig:cad}), which was assembled first. Due to this difference in the vertical position of the V-groove half-slits during fiber insertion, the left-routed fibers feeding the right V-groove block were inadvertently made slightly shorter since they were visually referenced to the raised left V-groove block, which made the apparent curvature righter. This avoidable occurrence led to tighter bending in the routing for the right V-groove block fibers compared to the left V-groove block. This can be discerned with careful inspection of Figure \ref{fig:routing}, panel (c). The effect of this slight difference in bending radius is described in detail later in \S~\ref{sec:performance}.

\subsection{Slit-end Polishing}
\label{sec:slit}

\begin{figure}[ht]
    \centering
    \includegraphics[width=0.85\linewidth]{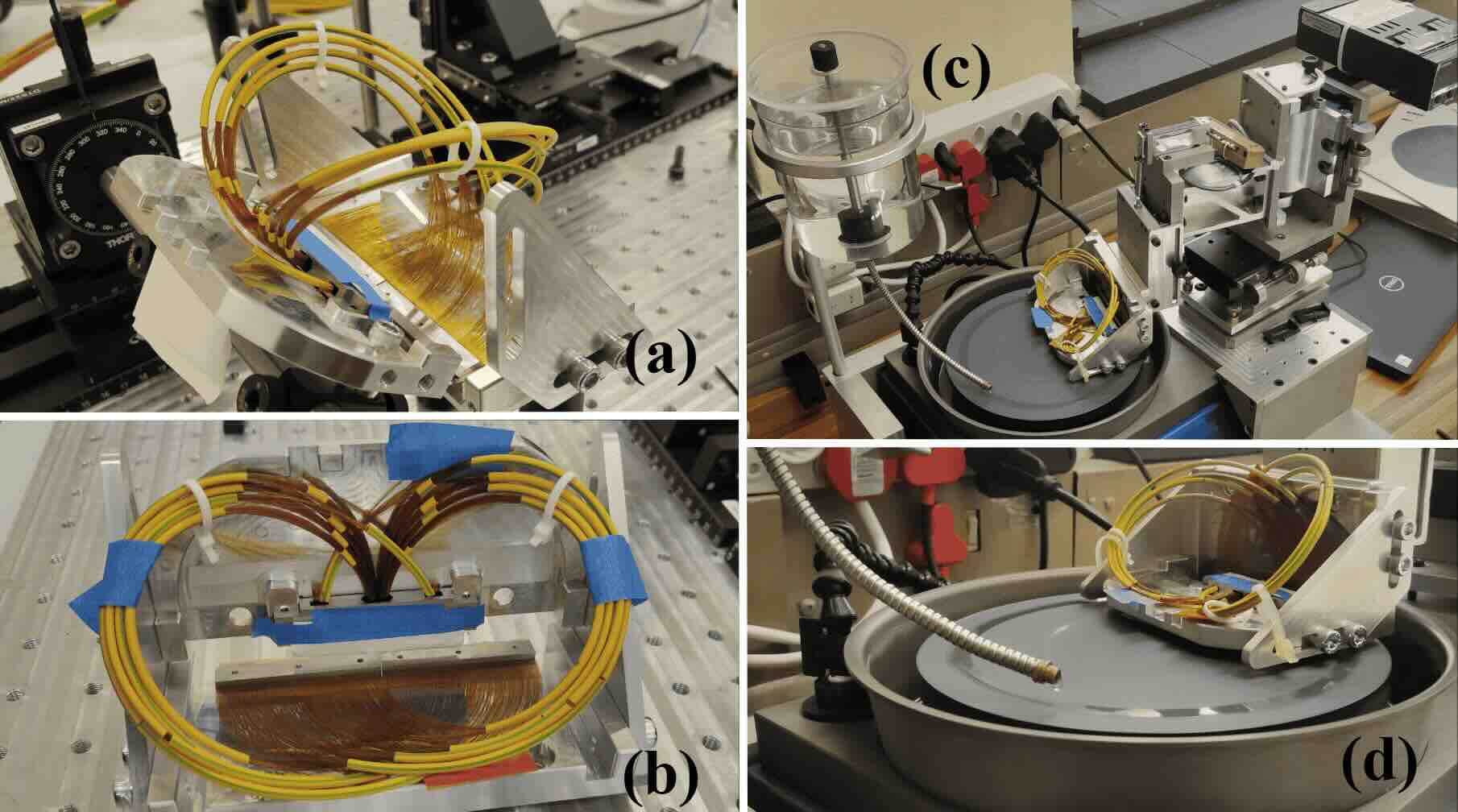}
    \caption{V-groove polishing process. Left panels (\textbf{a}) and (\textbf{b}) demonstrate the process of folding fiber assembly into a right-angled orientation between IFU block and slit blocks in order to access the V-groove assembly for polishing. Right panels (\textbf{c}) and (\textbf{d}) show the fiber assembly mounted on the polisher. }
    \label{fig:folding}
\end{figure}

Completion of rough slit termination was followed by slit polishing. To access the slit end of the fibers for polishing, the routed fiber assembly was folded into a right-angled configuration, as shown in Figure \ref{fig:folding}, panels (a) and (b). The entire assembly has two sections: the IFU side assembly and the slit side assembly. The IFU side assembly was mounted on a post attached to a rotating stage, with the rotation axis aligned to the post length. The rotating stage was mounted on a three-axis linear motion manual stage, while the slit side assembly was typically fixed on a post. Using the rotation and three-axis stages, the IFU side assembly was slowly rotated to achieve a right angle between the two assemblies. During this process, the fibers were lightly constrained using cable ties to prevent sudden motions. Once the right-angle configuration was achieved, two side plates were used to fix the configuration without the support of the rotation and linear stages, as shown in Figure \ref{fig:folding}, panels (a) and (b). This folded assembly was then mounted on the polisher and polished to a $1 \ \upmu\text{m}$ surface roughness, as shown in Figure \ref{fig:folding}, panels (c) and (d). Each of the left and right V-grooves was polished one at a time to the desired surface roughness. Slots were used to ensure the desired location of the fiber end surface with respect to the polishing surface. The fine termination of the slit was performed by following the same polishing recipe as described in \S~\ref{sec:skypol} for the IFU, except we modified the load in each grit configuration to 400 gm to account for the larger surface area. Given the large area of polishing (52.8 mm $\times$ 2 mm) of individual half slit blocks, we found that the polishing was less controlled. A smaller sub-slit (e.g. four quarter slit instead of two half slits) would lead to more deterministic surface finish on a shorter time scale.

With slit polishing complete, the assembly was then unfolded by reversing the folding process described above. The entire assembly nest of fibers, apertures, and V-grooves was transferred to the SMI-200 cassette using a transfer tool. The tool was designed to be attached to the polishing mount holes of fiber arrays and connect with the routed fibers with string ties. Once the fiber assembly was attached to the transfer tool, it was then detached from the routing assembly and transferred to the final cassette. We then mounted the fiber assembly on the SMI-200 cassette base and removed the transfer tool. 

\subsection{Final Assembly}
\label{sec:fass}

\begin{figure}[ht]
    \centering
    \includegraphics[width=0.5\linewidth]{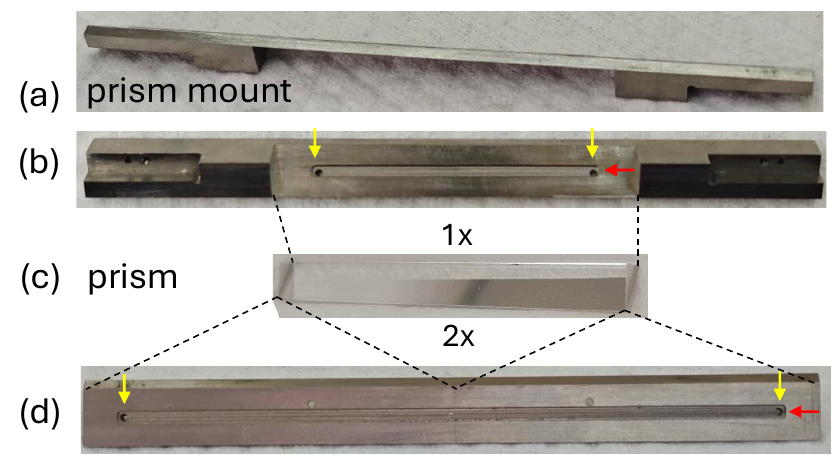}
 \caption{Prism mount viewed (a) edge-on; (b) from the bottom, showing the aperture for the IFU prism, the glue channel (horizontal red arrow), and insertion holes (vertical yellow arrows); (c) a Plexiglass mock prism, to scale; and (d) from the top, showing the glue channel and insertion holes.}
 \label{fig:prism1}
\end{figure}

In the final assembly phase, the IFU and fold prisms were slotted between the bi-dimensional (IFU) and linear (pseudo-slit) fiber arrays. All three prisms were mounted together on a custom fixture that produces precise separation between the slit and IFU prisms. The prism mount has several glue slots with glue injection and air escape holes as shown in Figure~\ref{fig:prism1}. 

\begin{figure}[ht]
    \centering
    \includegraphics[width=0.7\linewidth]{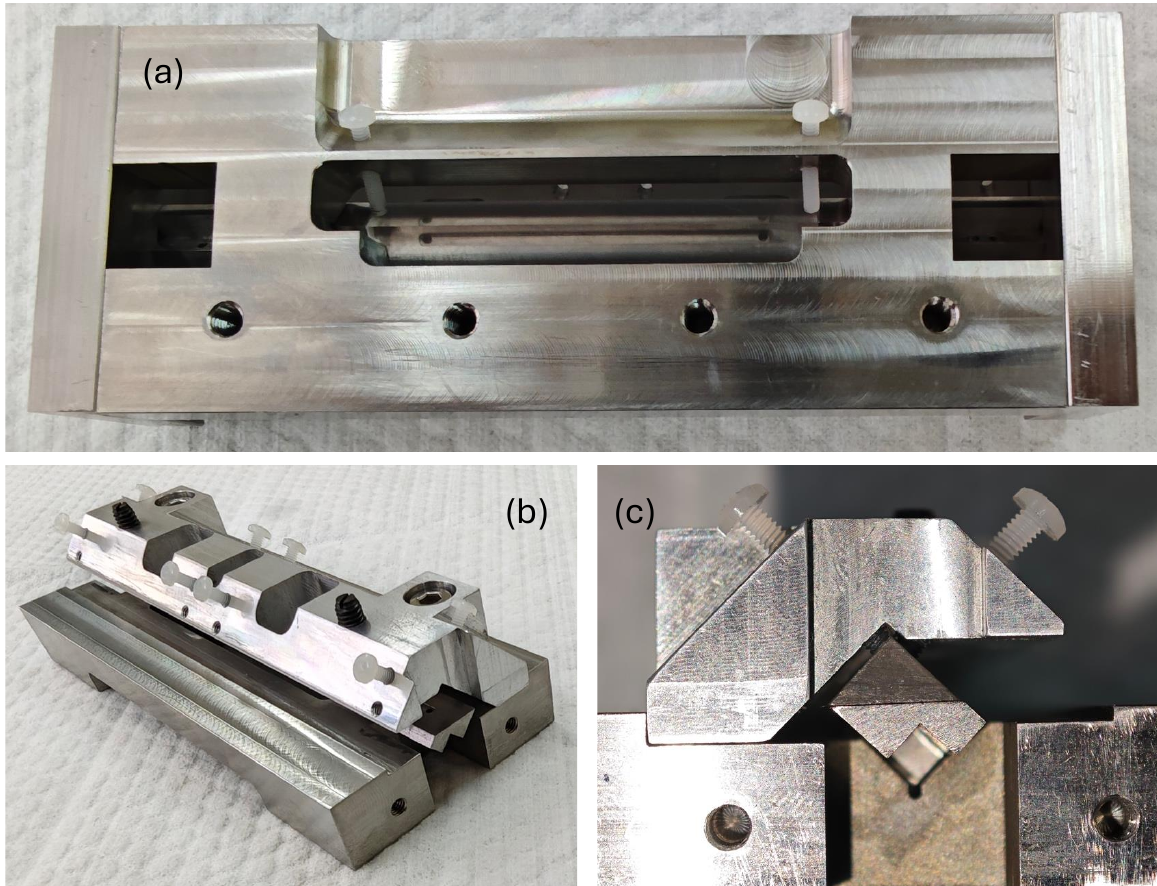}
 \caption{Prism assembly jig viewed (a) from the bottom (with side walls), showing the IFU prism and nylon positioning bolts; (b) from the top (without side walls), showing the prism mount without prisms; and (c) from the side, without side walls, showing the prism positioning on both top and bottom.}
 \label{fig:prism2}
\end{figure}

The prism assembly jig (Figure~\ref{fig:prism2}) has three parts: the base, the side blocks, and the lid. The base was designed to align and bond the IFU prism with the slot in the prism mount. The IFU prism was placed first on the assembly base, and the prism mount was dropped on it. There are two M2 nylon screws that adjust the position of the IFU prism by pushing in at 45 degrees from each side of the IFU prism. These screws are softer than metal screws and were positioned to contact the IFU prism outside the optical clear aperture. Spring-loaded, ball-head M3 bolts from the prism assembly lid were pushed down to keep the mount in place. Uncured Epotek-301 glue was injected through the insertion hole accessed from the top through the aperture on the assembly lid. The precuring process was avoided in this stage to ensure smooth filling of glue slot using injection syringe. The lateral movement of the mount was constrained by positioning side blocks on the two sides of the prism assembly. Spring-loaded, ball-head M3 bolts were used to keep the prism mount at the design horizontal position. Once the IFU prism bond was cured, the M3 bolts in the lid were removed, and the slit prisms were inserted after removing the side blocks. There were eight M2 nylon bolts holding the slit prism by pressing the prisms outside of their clear aperture and ensuring alignment. The assembly was completed by adding the side blocks. The bottom side of the base has aperture access to the slit side glue slot through which glue was poured in to bond the slit prisms. Before each glue application, alignment was checked under a microscope. Glue was injected slowly in order to avoid forming bubbles inside the slot. 

\begin{figure}[ht]
    \centering
    \includegraphics[width=\linewidth]{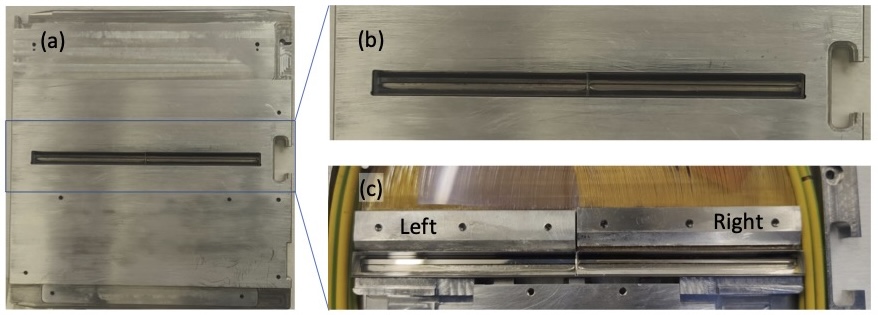}
    \caption{Left panel (\textbf{a}) shows the spectrograph facing side of the fully assembled SMI-200. The top-right panel (\textbf{b}) is a zoom-in view of the spectrograph opening, and the bottom-right panel (\textbf{c}) shows the positioning of the prism assembly under the lid between the slit and IFU. This figure is a modified version of figure 3 in  \cite{chattopadhyay2024}.}
    \label{fig:assembly}
\end{figure}

\begin{figure}[ht]
    \centering
    \includegraphics[width=0.7\linewidth]{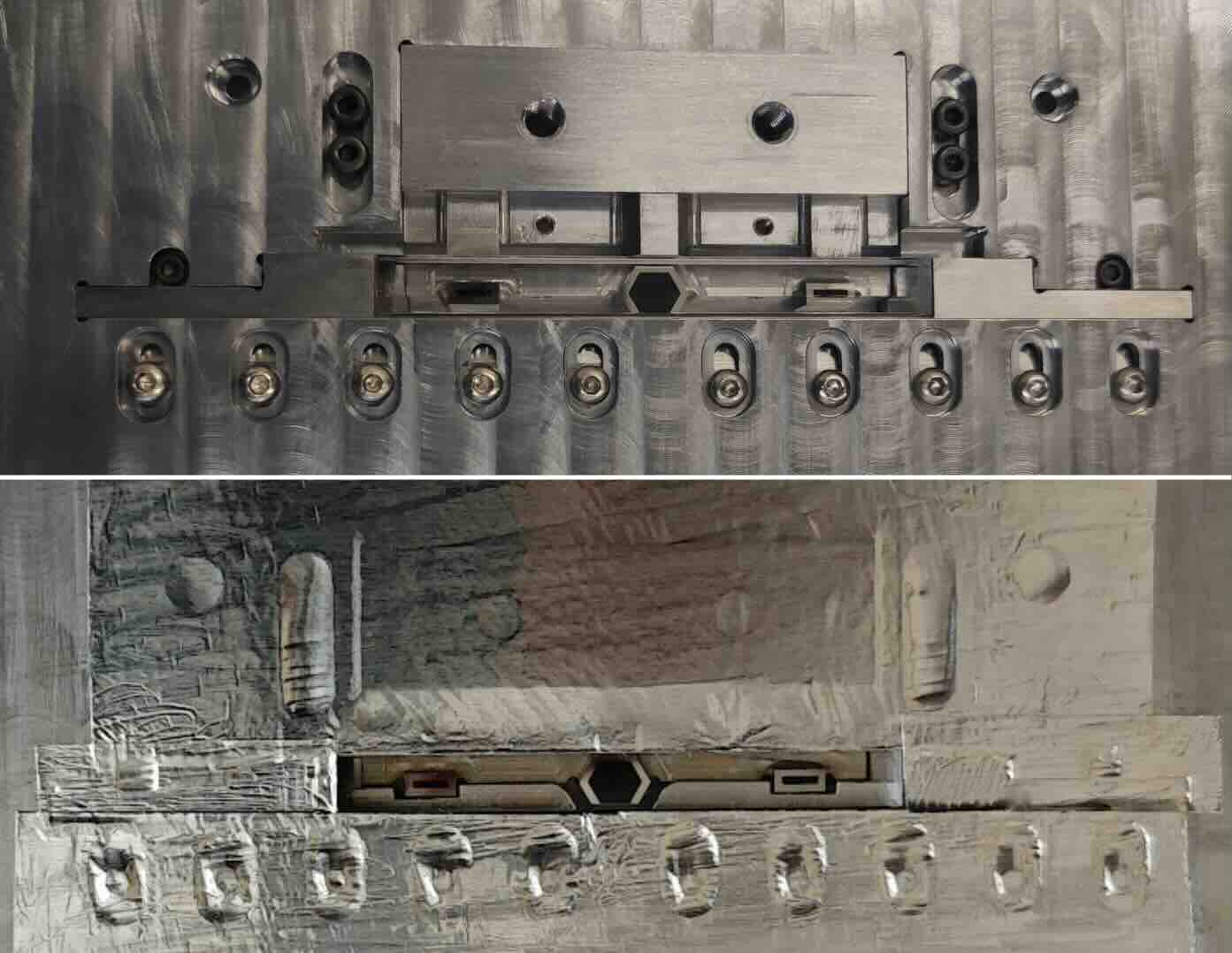}
    \caption{Stray light mitigation of the SMI-200 cassette. The top and bottom panels show before and after application of aluminum tape (on the telescope facing side of the cassette).}
    \label{fig:stray}
\end{figure}

Once cured, the prism assembly was transferred to the SMI-200 cassette base and held with two M1.6 bolts on the two sides of the IFU prism. The fiber ends were then pushed against the respective prism surfaces. Finally, the lids were attached to the cassette using M2 bolts, as shown in Figure \ref{fig:assembly}. Stray light through the SMI-200 cassette is mitigated with aluminum tape around the aperture and prism blocks at the telescope facing side of the cassette, as shown in figure \ref{fig:stray}. As discussed in Paper II, scattered light or ghosts arising from this tape or the steel cassette surface have not been identified, but should they be, this will be mitigated with flocking paint or other suitable blackening.

\section{Laboratory Characterization}
\label{sec:stray}

\subsection{SAAO-Wisconsin Fiber Testbed - SWiFT}
\label{sec:SWiFT}

An off-the-shelf component based custom fiber test stand (SAAO-Wisconsin Fiber Testbed - SWiFT) is used to test the throughput, FRD, and define the mapping of fibers between the sky and spectrograph end. The system is based on 25~mm optics in a cage-mounted assembly that is relatively low cost ($<20$k USD, base-date 2021, half of which is in stages and motors described in \S~\ref{sec:mapping}), compact (taking up less than 300~cm$\times$300~cm), and robust. A components list is provided in Appendix~\ref{app:swcomp} for reproduction.

\begin{figure}[ht]
    \centering
    \includegraphics[width=0.9\linewidth]{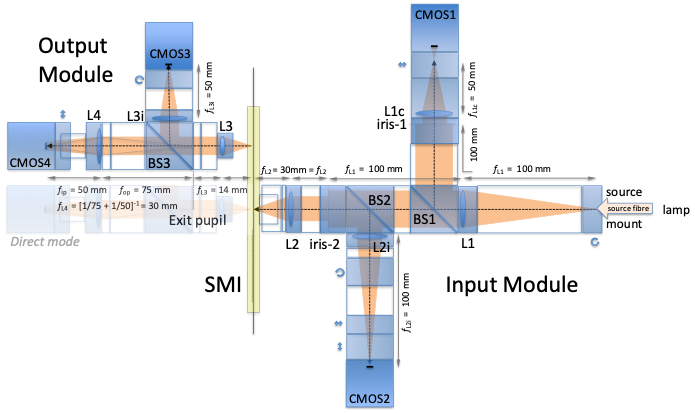}
    \caption{Schematic of the SAAO-Wisconsin Fiber Testbed (SWiFT), configured to test an SMI unit (vertical, yellow rectangle). The fiber injection module (Input Module), shown to the right, controls the input beam alignment and focal ratio with respect to the fiber being interrogated. The fiber output capturing module (Output Module), shown to the left, images both the fiber output near and far field energy distributions. The red vertical line bisecting the SMI unit represents the common focal plane of the input and output modules. Rotation, translation, and focus stages are indicated by corresponding arrows next to coupling elements. A vertically offset, transparent Output Module indicates its position for the ``Direct mode" used for calibration when the SMI unit is removed; actuation is described in the text. Critical focal lengths are labeled. Components are given in Table~\ref{tab:swcomp} of Appendix~\ref{app:swcomp}, referenced to labeled sub-assemblies. An important design element not captured in this schematic is the 90$^\circ$ rotation of the second beam-splitter (BS2) and the optical arm terminating in CMOS2 with respect to the first beam-splitter (BS1); this eliminates input beam polarization.}
    \label{fig:tester}
\end{figure}

SWiFT (Figure \ref{fig:tester}) uses a double-differential imaging system to determine how a fiber modulates an injected beam, with design origins in the fiber evaluation systems developed by \cite{Ramsey1988,Bershady_2004,drory}. Advantages of SWiFT over these earlier systems include the ability to determine fiber modulation in both the near- and far-field, while combining the automation capabilities of the systems described in \cite{drory} with the beam-speed, injection spot, and filtering flexibility of the test-stand described in \cite{Crause_2008}. Independently actuated Input and Output Modules straddle an intermediate focus where fiber optics is interrogated. The Input Module assembly of SWiFT ensures that the test fiber is fed a telecentric beam with the focal ratio equivalent to that of the telescope. In `fiber mode' operation, the output assembly captures both near and far-field intensity distribution of the fiber (fiber mode operation). In `direct mode' operation, the output assembly is aligned with the input assembly to measure the injected beam. Comparison of surface-photometry from output beam images between direct and fiber modes provides fiber performance quantification.

\subsubsection{SWiFT Input Module}
\label{sec:imod}

The Input Module uses a stabilized lamp (with an internal filter) coupled with a fiber SMA patch-cord as a source input to a collimating lens L1 (100~mm focal length spherical achromat). Placement of the filter inside the lamp, prior to the fiber coupling, eliminates significant internal reflections within the input system. A pellicle beam splitter (BS1) is placed in between the source and L1 to reflect 8\% of the beam onto CMOS1 (a low-cost, 1~Mpix array), which records and monitors the source intensity in a defocused image produced via lens L1c (a 50~mm focal-length aspherized achromat). A second pellicle beam splitter (BS2) is placed \textit{before} the pupil formed by L1. BS2 is oriented to reflect 8\% of the light returning (reflected) from the focal plane formed by L2 to a second CMOS device (CMOS2, identical to CMOS1). CMOS2 images the fiber input surface and injection spot via imaging lens L2i (100~mm focal length spherical achromat). Both beam-splitters transmit at 92\%, but are rotated 90$^\circ$ with respect to each other to eliminate instrumental polarization of the injection beam. This turns out to be important at the few~\% level for throughput calibration since fibers depolarize the injected light, and a third, identical beam-splitter in the output stage introduces modest polarization.

The transmitted beam in the Input Module is brought to an intermediate focus via a relatively fast lens L2 (focal length 30~mm), for which we use a aspherized achromat. The focal ratio is set by an adjustable iris (iris-2, (set to a diameter 7.143 mm for a f/4.2 injection beam) positioned \textit{at} the pupil formed by L1 and (for reflected light) L2. This ensures a crisp far-field injection beam profile. The spherical achromats introduce negligible spherical aberration, given they are used effectively at f/14. Overall, the optics deliver excellent on-axis image quality. The choice of 3.33:1 demagnification between the source and the intermediate foci is designed to allow for small input spots that sub-sampled the fibers under interrogation, given the multi-mode fiber patch-cords which are available with core ODs $>100~\upmu$m. For testing SMI-200, we used a $>105~\upmu$m core SMA with NA~0.28.

With this optical design using AR-coated optics and pellicle beam-splitters, the only significant internal reflection arises from the source fiber SMA ferrule, appearing in CMOS1. It was possible to mitigate this by adding a 4$^\circ$ tilt to the source mount plus a spatially offset iris (iris-1) in front of L1c.

\subsubsection{SWiFT Output Module}
\label{sec:pmod}

The Output Module accepts the input model injection (direct mode) or test fiber output (fiber mode) via lens L3 (a 12~mm diameter aspherized achromat with a focal length of 14~mm). A third pellicle beam-splitter (BS3) is placed behind the pupil formed by L3,  reflecting 8\% of the incoming beam to CMOS3 (identical to CMOS1/2), imaged via lens L3i (identical to L1c). CMOS3 captures the output near-field energy distribution by imaging the test fiber output surface. The pupil formed by L3 is re-imaged onto CMOS4 via lens L4 (identical to L2), capturing the far field from the fiber output or direct beam. The CMOS4 sensor (2~Mpix Sony IMX249), along with a 2/3 reduction in the pupil size provided by L4, yields an imaging field capable of capturing unvignetted output as fast as f/1.2.

\begin{figure}[ht]
    \centering
    \includegraphics[width=0.9\linewidth]{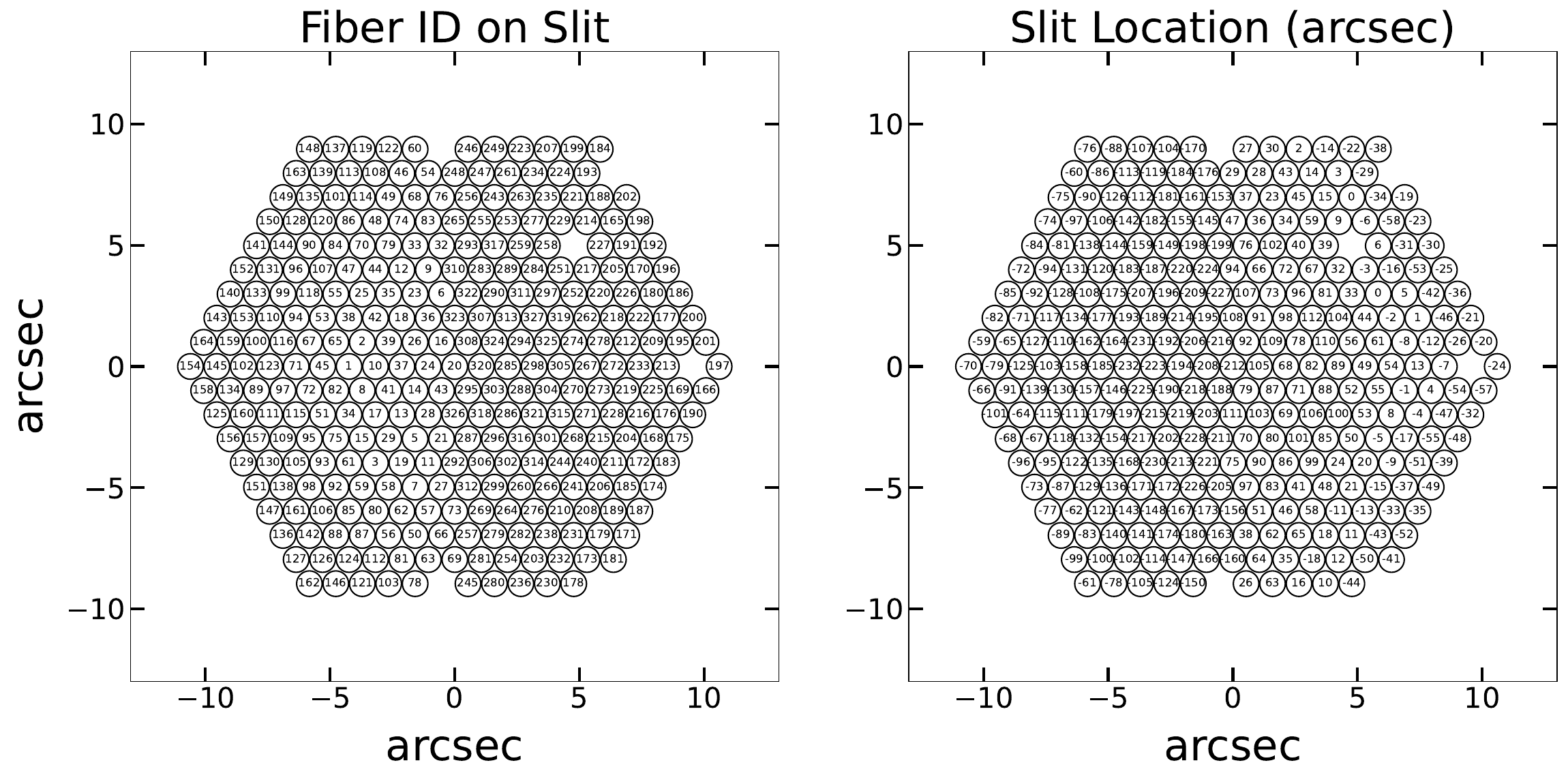}
    \caption{Fiber mapping on sky between the pseudo-slit and two-dimensional array. The left panel shows the fiber ID, where fiber ID 1 is at the bottom of the slit, while ID 327 is at the top. These numbers correspond to the slit number column in Appendix~\ref{app:fibermap}. The right panel shows the same, but the slit location is defined in arcsec, where 0 is the center of the slit, positive and negative values denote the right and left parts of the slit, respectively.}
    \label{fig:map}
\end{figure}

\subsection{Fiber Mapping: SWiFT actuation}
\label{sec:mapping}

The SWiFT Input Module is mounted on a three-axis linear stage. The optical axis stage (focus) is manually adjusted while the two orthogonal axes are motorized to enable computer-controlled mapping of fibers in the bi-dimensional fiber array of the IFU and sky bundles. The SWiFT output assembly is mounted on a motorized linear stage with the ability to scan along the slit dimension, but adjusted to the vertical position of the slit. The motorized coordinates of these three stages are registered using a LabVIEW script, which also controls the motion of the linear stages. 

Given the prism assembly does not change the location of the focal plane (as shown in Figure~\ref{fig:opp}), the insertion of the SMI-200 assembly does not require any focus adjustment; the separation between input and output assemblies is fixed between fiber and direct modes. By design, there is just room for the SMI cassettes to fit between the Input and Output modules at the common focal plane. The cassettes are held on a custom mount supported by two posts. Direct beam measurements are made once, before insertion of the SMI cassette, and then again after it is removed. Given lamp stability and monitoring via CMOS1, this is adequate for calibration.

At the input, individual fibers are illuminated using the Input Module, and a scan of the slit is performed via the Output Module. The initial fiber center was found manually on both input and output surfaces. Remaining fiber centers were then acquired via estimated spacing and motor calibration, but tweaked manually to refine the center positions. Based on the coordinates measured from the update centers, the mapping between sky and spectrograph ends of the IFU is defined, as shown in Figure~\ref{fig:map}, and tabulated in Appendix~\ref{app:fibermap} and electronically \href{https://drive.google.com/file/d/1SSrj73Y7_PaZKCeZCLHI2vlMIUoLEKdS/view?usp=sharing}{here}.

\subsection{Fiber Performance}
\label{sec:performance}

\subsubsection{Focal Ratio Degradation}
\label{sec:frd}

\begin{figure}[ht]
    \centering
    \includegraphics[width=\linewidth]{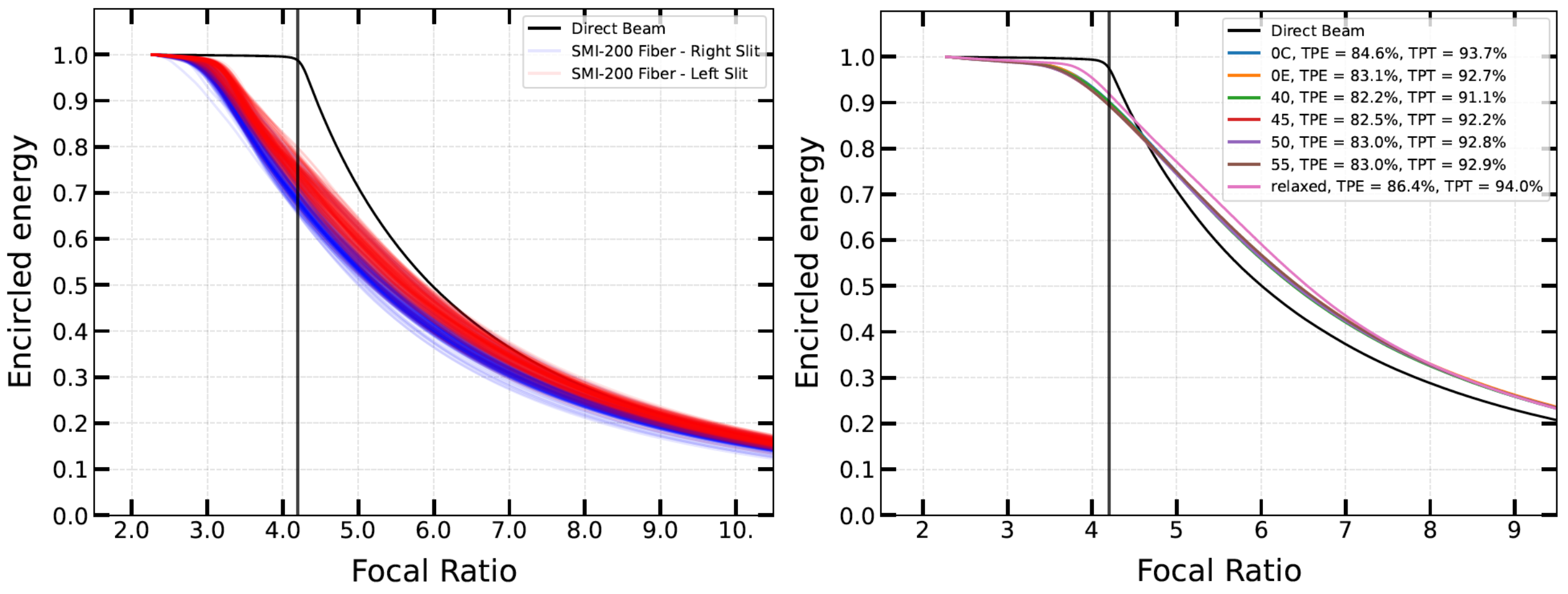}
    \caption{Normalized encircled energy of fiber output as a function of focal ratio. The left panel shows all the 327 fibers in SMI-200 (blue and red curves) with an injected direct f/4.2 beam (black curve). Blue and red encircled energy curves denote fibers in the right and left sides of the pseudo slit, respectively (see \S~{sec:tvar}). The median fiber output beam contains $\sim$70\% of the total output energy within the f/4.2 entrance beam of the RSS collimator. The right panel shows the performance of the calibration fibers when routed with curvatures mimicking SMI fibers at the slit center (0C) and slit edge (0E), as well as a range of bend diameters (mm), where the values of TPE and TPT correspond to effective and total throughput, respectively.}
    \label{fig:FRD}
\end{figure}

Figure \ref{fig:FRD} shows direct and fiber beam profiles of encircled energy vs focal ratio for SMI-200. The radius of 98\% encircled energy (EE98) is found to be at $\sim$f/2.5 for the fiber output beam when fed at f/4.2. However, RSS collimator optics are designed to accept an input of f/4.2. Within this acceptance cone, we find a median of $\sim$70\% of the total fiber output flux is captured. To understand the source of this FRD, we created and tested a single fiber reference cable (using the same 200:220:240 FBP fiber from Polymicro). The reference fiber was mounted similarly to the SMI-200 pseudo slit on a V-groove of similar length, attached to what would correspond to the center of the slit, with a routing with the tightest bending. The reference cable was tested using SWiFT and a f/4.2 injection beam. During the test, the reference cable was first held in a holder to mimic the bending radius of the SMI-200 fibers as routed from the IFU to the slit center (OC) and to the slit edge (OE). We also made measurements of the reference cable with even tighter bends, with diameters of 40 to 55 mm. The reference cable was also tested in a relaxed configuration where the smallest bending radius is 10 times larger than the bend radii in the SMI-200 routing configuration; we refer to this as the `relaxed' configuration.  As shown in Figure~\ref{fig:FRD}, the reference cable output captures about 90\%  of the light within the f/4.2 aperture regardless of configuration. Thus, we deduce that the large FRD exhibited by the SMI-200 fibers cannot be due primarily to the tight bending of fibers within the cassette. We surmise the elevated FRD in the SMI fibers can be attributed to the detailed routing of the fibers within the SMI-200 cassette, where fibers may get bunched up and twisted over sharp contours on mounting surfaces. We explore this in detail below.

\subsubsection{Total Throughput}
\label{sec:abstp}

\begin{figure}[ht]
    \centering
    \includegraphics[width=\linewidth]{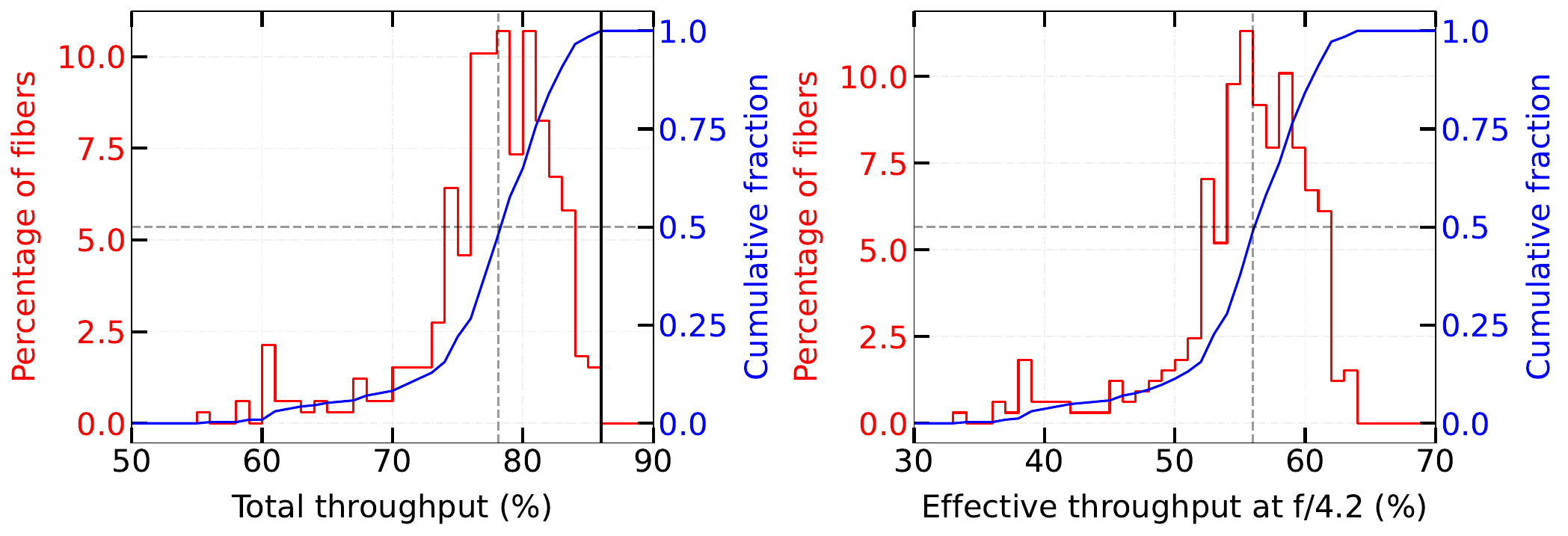}
    \caption{Histograms of total and effective fiber throughput for SMI-200, left and right panels, respectively. Effective throughput is measured within an aperture corresponding to a f/4.2 beam. The median total throughput is 77\%, while the median effective throughput is 56\%. The black vertical line at 87\% in the total throughput histogram represents the expected maximum value given SMI surface losses. This figure is a modified version of figure 5 in  \cite{chattopadhyay2024}.}
    \label{fig:tp_hist}
\end{figure}

The total throughput is defined using an aperture with a radius that fits entirely within the smaller detector dimension. In our case, this is f/1.182, which is significantly faster than the fiber numerical aperture. This still leaves ample detector for estimating any background levels from stray light. The fiber throughput is measured using a narrow-band red filter (655-685 nm) to provide a well-defined reference for on-telescope measurements. The median total throughput of the SMI-200 system is found to be 79.4\% over all fibers. We measured the total throughput of the reference cable to be $\sim$94\%, consistent with Fresnel losses (3.6\%) on the fiber entry and exit surfaces and negligible attenuation in the short fiber length. The total throughput of the SMI-200 system can be compared by including an additional four $\sim$1\% losses from the prism catheti, and two 97\% reflection losses (at 670~nm) from the protected-Ag prism folds. These combine to yield an expected 85\% maximum achievable total throughput. The maximum total throughput of SMI-200 fibers is measured to be $\sim$86\%, as shown in figure \ref{fig:tp_hist}. There is, however, significant variation in total throughput of SMI-200 fibers, likelystemming from mounting and routing within the assembly in ways significantly different from the reference fiber. The median total throughput is 77\%. Of the 336 fibers in SMI-200, we found three sky and six IFU fibers had no measurable throughput, despite no breakage (as indicated by inspection under a high-magnification microscope). These fibers are marked in Figure~\ref{fig:ntdist}. Hence, 24 sky and 303 object fibers ($>97$\%) are found to be usable.

\subsubsection{Effective Throughput}
\label{sec:reltp}

The effective throughput is defined as the amount of light fed by the SMI-200 fibers within the acceptable clear aperture diameter set by the spectrograph optics. Nominally, this corresponds to f/4.2 for RSS, although we defer the impact of NT until Paper~II. Combining the median total throughput (78\%) and the median encircled energy at f/4.2 (70\%), we expect the median effective throughput to be around $\sim$55\%.  Figure \ref{fig:tp_hist} shows that the median effective throughput of the SMI-200 fibers is 56\%, which is very close to our estimate. Indeed, the similar shapes of the total and effective throughput distributions reflect the rather narrow range of encircled energy profiles for the SMI-200 fibers.

\begin{figure}[ht]
    \centering
    \includegraphics[width=\linewidth]{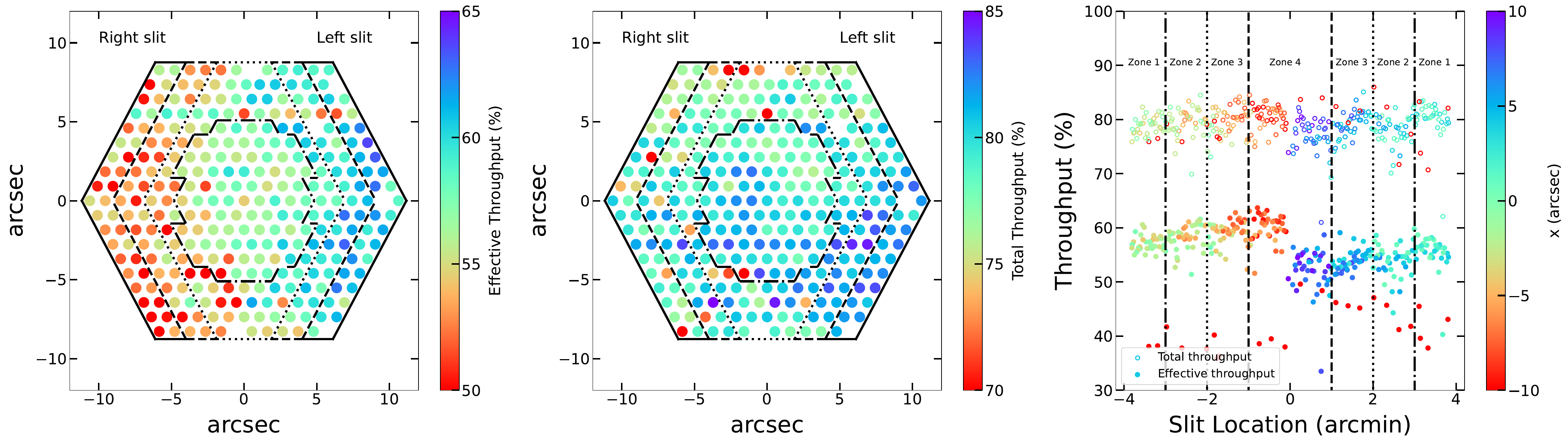}
    \caption{Spatial distribution of SMI-200 throughput within f/4.2 aperture (effective, left panel), and total throughput (middle panel) on sky and across the slit (right panel). The boundaries drawn represent the same non-telecentricity zones defined in figure \ref{fig:ntdist}. Variations may be attributed to the difference in the bending radius of left and right routed fibers. Bottom right panel shows the variation in effective throughput directly correlates to the fraction of encircled energy enclosed within the f/4.2 RSS clear aperture. }
    \label{fig:labtp}
\end{figure}

\subsubsection{Throughput Variation}
\label{sec:tvar}

The spatial variation of fiber throughput across the object aperture is shown in Figure~\ref{fig:labtp}, some -- but not all -- of which can be inferred from the color-coding in Figure~\ref{fig:FRD}. Fibers in the left and right sides of the IFU have different median effective throughput, with median values of 59\% and 54\%, respectively, for fibers mapped to the right and left slit blocks. This difference can be attributed to the tighter bending radius for fibers routed to the right-hand slit, as described in \S~\ref{sec:routing} and shown in the bottom-left panel of the figure \ref{fig:labtp}. The slight difference in bending radius is not by design and was introduced due to an inadvertent reference during fabrication (\S~\ref{sec:routing}) that will be corrected in subsequent SMI builds. However, there is no indication that total throughput varies from one side of the slit to the other which suggests that the increased FRD in most cases does not induce lossy modes.

Further inspection of the right panel of the figure \ref{fig:labtp} reveals modest trends in effective throughput with slit position. On average, a fiber sitting at the edge of the slit has roughly a 50\% larger bending radius on the slit side of the cassette compared to a fiber sitting closer to the slit center. This should introduce more FRD and hence a lower EE(f/4.2) and lower effective throughput moving toward the slit center. This appears to be the case for the right side of the slit (blue points), but not the left, where in fact the trend appears to be opposite. This opposition and the lack of dependence of FRD on bend radius for the reference cable suggest additional factors are at play. For example, close inspection of the pseudo-slits reveals that the left slit fibers have excess glue application extending behind the V-groove block (due to capillary wicking). The excess glue length increases from the center to the edge, correlating with higher FRD and lower effective throughput in the same direction (decreasing from the center to the edge). This suggests the bonding length and not curvature differences dominate the small (few \%) trends in effective throughput with slit position for the left slit. 

We also considered if there were other significant differences between the routing paths for fibers in the left and right slit halves, and between object and sky fibers. For example, the side channels do not have the same cross-section (see Figure~\ref{fig:cad}): The right channel (for fibers mapping to the left slit-block) has a smaller cross-section compared to the left channel (for fibers mapping to the right slit block). The channel volumes, therefore, do not appear to be a contributing factor in the differential performance between left and right slit blocks. Indeed, we inspected that the left slit fiber bundle cross-section is smaller than its channel, and hence there is no pinching effect. A more likely culprit leading to FRD performance differences may be how the fibers route immediately coming  out of their sky-side apertures. The right slit object fibers appear to be close to the metal wall of the IFU aperture; this may have led to a hinging effect on the aperture wall corner or at the glue joint that has increased the FRD seen in the right slit fibers. Consistent with this picture is the low-effective throughput of the sky fibers. This correlates with the tight bends at the location where these fibers come out of their metal aperture, which may also have led to stress from hinging at their glue joint or aperture wall corner. These effects can be mitigated in future SMI builds. In so doing, we aim to better ascertain what has led to the elevated levels of FRD in some of the SMI-200 fibers.

\section{Summary}
\label{sec:sum}

\subsection{Comparison with other IFUs in 10m class telescopes}
\label{sec:comparison}

SMIs are an IFS capability well suited to nebular spectroscopy at visible wavelengths on a 10~m class telescope, bringing SALT capabilities in line with other telescopes  8 to 10~m in size: Gemini \citep[GMOS,][]{Allington2002}), Keck \citep[KCWI,][]{Morrissey2018}), VLT \citep[MUSE,][]{muse}), GTC \citep[MEGARA,][]{megara}, HET \citep[VIRUS,][]{Hill2021}) and Subaru \citep[FOCAS,][]{Ozaki2020}. With the exception of the lenslet coupling, the GMOS IFU is the most similar to the SMI as a fiber-based retrofit to an existing imaging spectrograph. A comparison of instrument parameters is provided in Table~\ref{tab:comparison}\footnote{Since adaptive optics (AO) is not available on SALT, we have not included the parameters for MUSE and KCWI AO feeds that yield much finer spatial resolution and smaller fields of view. Similarly, we have not included near-infrared IFS and the many other IFS systems optimized for extreme AO at infrared wavelengths over very small fields.}, with the notable absence of throughput, which is often difficult to estimate from the literature. We will provide a detailed estimate of the on-sky throughput of SMI-200 in Paper II. 

In Table~\ref{tab:comparison}, $\lambda\lambda$ refers to the available spectral coverage, not the simultaneous spectral coverage in a given grating configuration; $\theta$ is the spatial resolution element (not pixel scale); while $N_R$ and $N_\theta$ refer to the number of independent spectral and spatial resolution elements.  We further define two performance metrics that characterize the ability of to constrain the kinematics for dynamically cold, low surface-brightness systems, such as galaxy disks: (1) $\sqrt{A\!\times\!\Omega_\theta}\!\times\!R$, where $A$ is the telescope area ($\pi [D_T/2]^2]$), $\Omega_\theta$ is the solid angle of resolution element ($\pi [\theta/2]^2$), and $R$ is the spectral resolution of the telescope-instrument system; and (2) a figure of merit, $FoM = \sqrt{N_R\!\times\!N_\theta\!\times\!A\!\times\!\Omega_\theta}\! \times\!R$. The first metric is the product of the signal-to-noise and spectral resolution, which scales with, e.g., the precision in determining line centroids. The second metric combines the first with the product of the square-root of the number of sampling elements (spatial and spectral); this should scale with the ability to constrain kinematic models, e.g., a velocity field. With the FoM we are assuming that on average each spatial and spectral resolution element contributes comparable information. In the case of spectral elements, this assumption would be valid where the kinematic information is contained in many weak lines distributed across the observed band-pass. With the tabulated information one can also compute various general merit functions found in the literature \citep[e.g., Bershady's contribution in][]{3DSpec2009,Hill2014,Melso2022}, such as total grasp, $A \times N_\theta \times \Omega_\theta$, and spectral power, $N_R \times R$. We prefer the specific merit functions introduced here because of their direct relevance to instrument performance for the specific science for which the SMI instruments were designed.

\begin{deluxetable}{cccccccccc}
\label{tab:comparison}
\tabletypesize{\postscriptsize}
\tablewidth{90pt}
\tabletypesize{\scriptsize}
\centering
\tablecaption{Comparison of SMI with Wide-Field IFS on 10m class telescopes}
\tablehead{\colhead{Instrument} & \colhead{D$_T$} & \colhead{$\lambda\lambda$} & \colhead{FoV} & \colhead{$R$}  &
\colhead{$\theta$} & \colhead{$N_R$} & \colhead{$N_\theta$} & \colhead{$\sqrt{A\!\times\!\Omega_\theta}\!\times\!R$}  & \colhead{$FoM$} \\ [-0.1in]
\colhead{/ Telescope} & \colhead{(m)} & \colhead{(nm)} & \colhead{(arcsec$\times$arcsec)} & \colhead{($\lambda/\delta\lambda$)} & \colhead{(arcsec)} & \colhead{} & \colhead{} & \colhead{(mm~\SI{}{\radian})} & \colhead{(10$^4$ mm~\SI{}{\radian})}} 
\startdata
GMOS$^a$ / Gemini  & 8.1  & 360-1030 & $5\times(7+3.5)$   &  500-8800  & 0.2   &  500 &  1500 &  3.1-54.3 & 0.27-4.7 \\
FOCAS / Subaru & 8.2  & 370-995  & $13.5\times10$         &  900-2700  & 0.435 &  600 &   713 & 12.2-36.7 & 0.80-2.4 \\
MUSE-WFM / VLT & 8.2  & 465-930  & $60\times60$           & 3000       & 0.35  & 2000 & 29388 &      32.8 & 25 \\
VIRUS / HET & 10  & 350-550  & $78\times(51\times51)$ &  800       & 1.5   &  356 & 34944 &      43.8 & 15.6 \\
KCWI-S$^b$ / Keck  & 10   & 350-1080 & $20\times8$        & 3600-18000 & 0.35  & 2640 &  1306 &  48.0-240 & 8.9-44 \\
KCWI-M$^b$ / Keck  & 10   & 350-1080 & $20\times16$       & 1800-9000  & 0.70  & 1320 &   653 &  48.0-240 & 4.5-22 \\
KCWI-L$^b$ / Keck  & 10   & 350-1080 & $20\times33$       &  900-4500  & 1.39  & 660  &   342 &  47.6-238 & 2.3-11 \\
MEGARA / GTC   & 10.4 & 365-888  & $12.5\times11.3$       & 5500-20000 & 0.62  & 1100 &   367 &   135-491 & 8.6-31\\
{\bf SMI-200 / SALT}&\textbf{9.2}&\textbf{320-900}&\textbf{$22.5\times17.6$}&\textbf{800-9000}&\textbf{0.88}& 1060 &\textbf{336}&\textbf{24.7-277}&\textbf{1.5-17}\\
SMI-300 / SALT & 9.2  & 320-900  & $33\times18.6$         & 530-6000   & 1.33  &  700 &   250 &  24.7-280 & 1.0-12\\
SMI-400 / SALT & 9.2  & 320-900  & $40.6\times20.5$       & 400-4500   & 1.77  &  526 &   200 &  24.8-279 & 0.80-9.0\\
\bottomrule
\enddata
\tablenotetext{a}{GMOS 2-slit mode including both object and sky arrays.}
\tablenotetext{b}{Both red and blue KCWI channels.}
\end{deluxetable}

Most of the IFS instruments in Table~\ref{tab:comparison} have (order of magnitude) 10$^3$ each of spatial and spectral resolution elements, within factors of a few. VIRUS and MUSE are exceptions, but only in terms of $N_\theta$ due to their large number of spectrographs; in contrast, their number of spectral elements ($N_R$) is more typical, and their
$\sqrt{A\!\times\!\Omega_\theta}\!\times\!R$ values are more modest, reflecting the capabilities of the individual spectrographs. GMOS, FOCAS, and VIRUS put more of a premium on spatial information relative to spectral information, while KCWI, MEGARA, and SMI all have significantly more simultaneous spectral coverage and/or spectral resolution; MUSE achieves both but with reduced $\sqrt{A\!\times\!\Omega_\theta}\!\times\!R$. In the context of our metrics, SMI is very similar to KCWI and MEGARA.  Despite the significant advantages the other, more optimized instruments provide, SMIs represent a low-cost fiber-optic retrofit to an existing spectrograph that still yields impressive performance capabilities. 

\subsection{Lessons learned}
\label{sec:lessons}

The SMI instrument series aims to provide large, two-dimensional, on-sky footprints by reformatting as much of the 8$\arcmin$ RSS slit using fibers packed as tightly as practical given cross-talk constraints (\S~\ref{sec:crosstalk}). In practice, with SMI-200, we were able to use 101 mm of the 105.6 mm slit-length, limited only by mechanical walls and gaps of the V-groove assemblies comprising the pseudo-slit assembly. The fiber-to-fiber separation for SMI-200 yields a total fiber count of 336, while delivering 1:10 contrast or better between fiber traces in all possible RSS grating and camera configurations for an idealized system. In Paper II these cross-talk predictions will be compared to the as-built performance of the spectrograph that includes as-built optical misalignment and scattered light.

Given the thickness constraints of the SMI cassette and the requirements for par-focality of the SMI system with long-slit and MOS cassettes, we determined that the maximum height of the IFUs is limited to 5.59 mm, of which we used 3.98 mm in the SMI-200 design. The resulting extended hexagonal shapes provide a sufficient match to the typical elongation of galaxies on the sky.

The severe NT of the SALT focal plane proved to be manageable for both IFU and fiber pseudo-slit fabrication. At the slit end, positioning the fibers with a continuously increasing NT angle from the center to the edge of the slit proved to be straightforward with standard wire-EDM techniques used in fabricating the V-grooves. Although the fabricated angles were inverted in this first SMI-200 assembly, this will be corrected in future builds. We found that plunge-EDM techniques for fabricating the IFU mounts were also sufficient to tilt the sky arrays to compensate for the on-telecentricity of the SALT focal plane. This enabled us to provide simultaneous observation of the sky foreground along with the object to eliminate sky subtraction issues due to the variability in the telescope pupil between different tracks of observation. 

Significant care in bonding agent (epoxy) pre-curing and its application has enabled a reduction in the number of gaps between fibers, thereby improving the quality of finish. However, the duration of the required fiber polishing to achieve a given surface finish was found to depend significantly on the area of the polished surface, despite not being limited by usable polishing disc size. The polished surface area includes both fibers and their mounts. A larger area led to significantly longer polishing times to eliminate surface scratches and dig. Although it is not possible to modify the termination process of the sky end (the IFU and sky arrays), future developments can reduce the surface area of the V-groove slit blocks by dividing the slit into four (quarter) blocks rather than the two half-slits of the current design. This modification to future SMI units will also (i) improve handling of large core fibers by reducing the spring force on a given slit block; and (ii) provide additional flexibility in routing fibers through the asymmetrically sized left and right channels.

During the assembly process, routing and termination of the fibers appear to have created differences in fiber performance between the left and right of the slit blocks, sky fibers, and along the slit. The right slit fibers were found to have tighter bend radii compared to the left slit fibers, but this may not be the only or even the leading cause of performance differences. The right slit object fibers and the sky fibers appear to be too close to the\textbf{ir} metal aperture wall, possibly causing a hinging effect on the wall corners or the glue joints, exacerbating stress-induced FRD. A more open aperture design, as well as better attention to the fiber routing coming out of these apertures, should improve performance.

The left slit fibers are found to have excess glue wicking, increasing from the center to the slit edge. This appears to explain trends of increasing FRD along this portion of the slit. Additional attention to the pre-cure process and glue volume application should mitigate this problem.

Fiber performance has been well characterized using SWiFT, a test-station based on commercial components. A calibration fiber demonstrates that the effect of fiber bend radius does not dominate throughput performance, supporting our inference that the intricate issues of fiber mounting apertures and glue wicking dominate the realized performance of SMI-200. While the effective throughput for SMI-200 is found to be 56\%, it should be easily possible to increase performance to 62\% and possibly as high as 78\% once all assembly artifacts are eliminated in future versions.

\section* {Acknowledgments} 
This research was supported by funds from the University of Wisconsin-Madison Department of Astronomy Board of Visitors and the South African National Research Foundation SARChI program. Figure \ref{fig:polish}, figure \ref{fig:assembly}, \ref{fig:tp_hist} of this paper is reproduced from figure 2, figure 3 and figure 5 of \cite{chattopadhyay2024} respectively.

\bibliography{report}{}
\bibliographystyle{aasjournalv7}

\appendix

\section{SWiFT Components}
\label{app:swcomp}

Table~\ref{tab:swcomp} contains a complete commercial parts list for SWiFT as described in \S~\ref{sec:SWiFT}, referencing sub-assemblies labeled in Figure~\ref{fig:tester} in addition to motion control components.

\centerwidetable 
\startlongtable
\begin{deluxetable}{llll}

\label{tab:swcomp}
\tablewidth{90pt}
\tabletypesize{\scriptsize}
\centering

\tablecaption{SWiFT components list.}

\tablehead{\colhead{\textbf{Component description}} & \colhead{\textbf{Sub-assembly}} & \colhead{\textbf{Vendor}} & \colhead{\textbf{Part}} } 

\startdata
Stabilized fiber-coupled Light Source  & Lamp & Thorlabs & 	SLS201L/M \\
SMA Fiber Patch Cable: 105 µm, 0.22 NA, 2m  & Lamp & Thorlabs & M15L02 \\
SMA Fiber Adapter Plate with External SM1 (1.035"-40) Threads & Source & Thorlabs & SM1SMA \\ 
SM1 (1.035"-40) Coupler, External Threads, 0.5" Long, Two Locking Rings & Source & Thorlabs & SM1T2 \\
Cage Assembly Rod, 4" Long, Ø6 mm, 4 Pack & Source & Thorlabs & ER4-P4 \\
Cage Assembly Rod, 1/2" Long, Ø6 mm, 4 Pack & Source & Thorlabs & ER05-P4 \\
SM1 Lens Tube, 1.00" Thread Depth, One Retaining Ring Included & Source & Thorlabs & SM1L10 \\
SM1 Lens Tube, 1.50" Thread Depth, One Retaining Ring Included & Source & Thorlabs & SM1L15 \\
Ø1" Adjustable Lens Tube, 1.31" Travel Range & Source & Thorlabs & SM1V15 \\
f=100 mm, Ø1" Achromatic Doublet, SM1-Threaded Mount, ARC: 400-700 nm & L1 & Thorlabs & AC254-100-A-ML \\
SM1 (1.035"-40) Coupler, External Threads, 0.5" Long, Two Locking Rings & L1 & Thorlabs & SM1T2 \\
Cube-Mounted Pellicle Beamsplitter, 8:92 (R:T), Uncoated, 400 - 2400 nm & BS1 & Thorlabs & CM1-BP108 \\
Externally SM1-Threaded End Cap & BS1 & Thorlabs & SM1CP2 \\
SM1 Lens Tube, 1.00" Thread Depth, One Retaining Ring Included & L1c+CMOS1 & Thorlabs & SM1L10 \\
SM1 Lens Tube, 2.00" Thread Depth, One Retaining Ring Included & L1c+CMOS1 & Thorlabs & SM1L20 \\
Cage Assembly Rod, 6" Long, Ø6 mm, 4 Pack & L1c+CMOS1 & Thorlabs & ER6-P4 \\
30 mm Cage System, XY Translating Lens Mount for Ø1" Optics & L1c+CMOS1 & Thorlabs & CXY1 \\
Adapter with External C-Mount Threads and External SM1 Threads, 3.2 mm Spacer & L1c+CMOS1 & Thorlabs & SM1A39 \\
CMOS, 1280x960, AR0130CS, 4e- RN, 3.75um pix, 12 bit, 375-775m QE50 & L1c+CMOS1	& ZWO& ASI120MM mini \\
25mm Diameter, 50mm EFL Aspherized Achromatic Lens & L1c+CMOS1	& Edmund Optics & 49-665 \\
Kinematic, SM1-Threaded, 30 mm-Cage-Compatible Mount with Slip Plate for Ø1" Optic & L2i+CMOS2	& Thorlabs & KC1-S/M \\
SM1 (1.035"-40) Coupler, External Threads, 0.5" Long, Two Locking Rings & L2i+CMOS2 & Thorlabs & SM1T2 \\
Cage Assembly Rod, 1.5" Long, Ø6 mm, 4 Pack & L2i+CMOS2 & Thorlabs & ER1.5-P4 \\
SM1 Lens Tube, 1.00" Thread Depth, One Retaining Ring Included & L2i+CMOS2 & Thorlabs & SM1L10 \\
SM1-Threaded 30 mm Cage Plate, 4.0 mm Thick & L2i+CMOS2 & Thorlabs & CP4S \\
f=100 mm, Ø1" Achromatic Doublet, SM1-Threaded Mount, ARC: 400-700 nm & L2i+CMOS2 & Thorlabs & AC254-100-A-ML \\
CMOS, 1280x960, AR0130CS, 4e- RN, 3.75um pix, 12 bit, 375-775m QE50 & L2i+CMOS2	& ZWO	& ASI120MM mini \\
Cube-Mounted Pellicle Beamsplitter, 8:92 (R:T), Uncoated, 400 - 2400 nm & BS2 & Thorlabs & CM1-BP108 \\
Externally SM1-Threaded End Cap & BS2 & Thorlabs & SM1CP2 \\
Cage Assembly Rod, 1/2" Long, Ø6 mm, 4 Pack & BS2 & Thorlabs & ER05-P4 \\
Rod Adapter for Ø6 mm ER Rods, 4 Pack & BS2 & Thorlabs & ERSCA-P4 \\
SM1 Lens Tube, 0.50" Thread Depth, One Retaining Ring Included & BS2 & Thorlabs & SM1L05 \\
Extra-Thick SM1 (1.035"-40) Threaded Retaining Ring & BS2 & Thorlabs & SM1RRC \\
SM1 Graduated Ring-Actuated Zero Aperture Iris Diaphragm, Ø12.0 mm Max Aperture & Iris & Thorlabs & SM1D12CZ \\
Cage Assembly Rod, 2" Long, Ø6 mm, 4 Pack & L2 & Thorlabs & ER2-P4 \\
Extra-Thick SM1 (1.035"-40) Threaded Retaining Ring & L2 & Thorlabs & SM1RRC \\
SM1 Lens Tube, 0.50" Thread Depth, One Retaining Ring Included ($\times$2) & L2 & Thorlabs & SM1L05 \\
SM1 Lens Tube, 0.30" Thread Depth, One Retaining Ring Included & L2 & Thorlabs & SM1L03 \\
SM1-Threaded 30 mm Cage Plate, 6 mm Thick & L2 & Thorlabs & CP6S/T \\
25mm Diameter, 30mm EFL Aspherized Achromatic Lens & L2 & Edmund Optics & 49-662 \\
SM05-Threaded 30 mm Cage Plate, 0.35" Thick, Two Retaining Rings, M4 Tap & L3 & Thorlabs & CP11/M \\
SM05 Lens Tube, 0.30" Thread Depth, One Retaining Ring Included & L3 & Thorlabs & SM05L03 \\
SM05 Lens Tube, 0.50" Thread Depth, One Retaining Ring Included & L3 & Thorlabs & SM05L05 \\
Cage Assembly Rod, 1/2" Long, Ø6 mm, 4 Pack & L3 & Thorlabs & ER05-P4 \\
SM1-Threaded 30 mm Cage Plate, 4.0 mm Thick & L3 & Thorlabs & CP4S \\
12.5mm Diameter, 14mm EFL Aspherized Achromatic Lens & L2 & Edmund Optics & 49-658 \\
Kinematic, SM1-Threaded, 30 mm-Cage-Compatible Mount with Slip Plate for Ø1" Optic & L3i+CMOS3 & Thorlabs & KC1-S/M  \\
Adapter with External C-Mount Threads and External SM1 Threads, 3.2 mm Spacer & L3i+CMOS3 & Thorlabs & SM1A39 \\
SM05 Lens Tube, 0.50" Thread Depth, One Retaining Ring Included ($\times$2) & L3i+CMOS3 & Thorlabs & SM05L05 \\
Adapter with External SM05 Threads and External SM1 Threads & L3i+CMOS3 & Thorlabs & SM05A3 \\
Cage Assembly Rod, 2" Long, Ø6 mm, 4 Pack & L3i+CMOS3 & Thorlabs & ER2-P4 \\
SM1 Lens Tube, 0.50" Thread Depth, One Retaining Ring Included & L3i+CMOS3 & Thorlabs & SM1L05 \\
CMOS, 1280x960, AR0130CS, 4e- RN, 3.75um pix, 12 bit, 375-775m QE50 & L3i+CMOS3	& ZWO	& ASI120MM mini \\
25mm Diameter, 50mm EFL Aspherized Achromatic Lens & L3i+CMOS3	& Edmund Optics & 49-665 \\
Cube-Mounted Pellicle Beamsplitter, 8:92 (R:T), Uncoated, 400 - 2400 nm & BS3 & Thorlabs & CM1-BP108 \\
Externally SM1-Threaded End Cap & BS3 & Thorlabs & SM1CP2 \\
SM1 Lens Tube, 0.50" Thread Depth, One Retaining Ring Included & L4+CMOS4 & Thorlabs & SM1L05 \\
SM1 Lens Tube, 1.00" Thread Depth, One Retaining Ring Included & L4+CMOS4 & Thorlabs & SM1L10 \\
Cage Assembly Rod, 3" Long, Ø6 mm, 4 Pack & L4+CMOS4 & Thorlabs & ER3-P4 \\
Adapter with External C-Mount Threads and External SM1 Threads, 3.2 mm Spacer & L4+CMOS4 & Thorlabs & SM1A39 \\
30 mm Cage System, XY Translating Lens Mount for Ø1" Optics & L4+CMOS4 & Thorlabs & CXY1 \\
Adapter with External SM05 Threads and External SM1 Threads & L4+CMOS4 & Thorlabs & SM05A3 \\
Adapter with External SM1 Threads and Internal SM05 Threads, 0.15" Long ($\times$2) & L4+CMOS4 & Thorlabs & SM1A6 \\
SM1-Threaded 30 mm Cage Plate, 8 mm Thick & L4+CMOS4 & Thorlabs & CP8S/T \\
1936x1216 CMOS  Sony IMX249, 3.5e- RN, 5.86um pix, 12 bit, 375-775m QE50 & L4+CMOS4 & ZWO & ASI174MM Mini \\
25mm Diameter, 30mm EFL Aspherized Achromatic Lens & L4+CMOS4 & Edmund Optics & 49-662 \\
150 mm Translation Stage with Stepper Motor, Integrated Controller, M6 Taps & Mounting & Thorlabs & LTS150/M \\
Self-Contained XYZ 25 mm Translation Stage, M6 x 1.0 Taps & Mounting & Thorlabs & LX30/M \\ 
25 mm Motorized Actuator with Ø3/8" Barrel (0.5 m Cable) ($\times$2) & Mounting & Thorlabs & Z825B \\
K-Cube Brushed DC Servo Motor Controller (Power Supply Not Included) ($\times$2) & Mounting & Thorlabs &  KDC101 \\
USB Controller Hub and Power Supply for Three K-Cubes or T-Cubes & Mounting & Thorlabs & KCH301 \\
DC Servo Motor Cable for Z9 Motors, DE15 Male to DE15 Female, 2.5 m ($\times$2) & Mounting & Thorlabs &  PAA632 \\
Right-Angle Bracket with Counterbored Slots and M6 Tapped Holes ($\times$2) & Mounting & Thorlabs &  AB90A/M \\
Dovetail Optical Rail, 75 mm & Mounting & Thorlabs & RLA075/M \\
Dovetail Optical Rail, 150 mm, Metric & Mounting & Thorlabs &  RLA150/M \\
Dovetail Rail Carrier, 1.00" x 1.00" (25.4 mm x 25.4 mm), 1/4" (M6) Counterbores ($\times$2) & Mounting & Thorlabs & RC1 \\
\bottomrule
\enddata
\end{deluxetable}

\section{Astrometry}
\label{app:fibermap}

Table~\ref{tab:astrometry} contains the mapping of fibers between the RSS pseudo-slit and the IFU and sky arrays, as well as the laboratory-measured throughput described in \S~\ref{sec:performance}. Table columns have the following definitions. (a) Slit: trace location of the fiber from the bottom of the RSS detector. (b) X: Offset location of the fiber on sky in arcsec from the IFU array center defined by the fiber with a Slit ID of 20 along the x direction of the observed position angle. (c) Y: Offset location of the fiber on sky in arcsec from the IFU array center defined by the fiber with a Slit ID of 20 along the y direction of the observed position angle, (d) ET: Effective throughput of the SMI-200 fibers within f/4.2 aperture of SALT-RSS. (e) TT: Total throughput of the SMI-200 fibers. (f)Tp: Designation of the fiber in the object IFU (o) or one of the two sky arrays (s).

\startlongtable
\begin{deluxetable}{cccccccccccccccccccccccccccc}
\setlength{\tabcolsep}{1pt}
\label{tab:astrometry}
\tabletypesize{\scriptsize}
\centering

\tablecaption{SMI-200 fiber mapping and in-lab performance.}

\tablehead{\colhead{\textbf{Component description}} & \colhead{\textbf{Sub-assembly}} & \colhead{\textbf{Vendor}} & \colhead{\textbf{Part}} } 

\tablehead{\colhead{} & \colhead{} & \colhead{\textbf{Right Slit}} & \colhead{} & \colhead{} & \colhead{} & \colhead{} & \colhead{} & \colhead{\textbf{Right Slit}} & \colhead{} & \colhead{} & \colhead{} & \colhead{} & \colhead{} &  \colhead{\textbf{Left Slit}} & \colhead{} & \colhead{} & \colhead{} & \colhead{} & \colhead{} & \colhead{\textbf{Left Slit}} & \colhead{} & \colhead{} & \colhead{} & \colhead{} &  \\
\textbf{Slit} & \textbf{X} & \textbf{Y} & \textbf{ET}   & \textbf{TT}  & \textbf{Tp} & \textbf{Slit} & \textbf{X} & \textbf{Y} & \textbf{ET}   & \textbf{TT}  & \textbf{Tp} & \textbf{Slit} & \textbf{X} & \textbf{Y} & \textbf{ET}   & \textbf{TT}  & \textbf{Tp} & \textbf{Slit} & \textbf{X} & \textbf{Y} &  \textbf{ET}   & \textbf{TT}  & \textbf{Tp} \\ 
(a) & (b) & (c) & (d) & (e) & (f) & (a) & (b) & (c) & (d) & (e) & (f) & (a) & (b) & (c) & (d) & (e) & (f) & (a) & (b) & (c) & (d) & (e) & (f) }

\startdata
\textbf{1} & -4.345 & -0.04 & 54.6 & 78.9 & o & \textbf{83} & -1.035 & 5.673 & 56.9 & 75.8 & o & \textbf{165} & 6.513 & 5.633 & 51.6 & 74.3 & o & \textbf{247} & 1.075 & 7.686 & 60.6 & 78.3 & o\\
\textbf{2} & -3.827 & 0.841 & 43.1 & 80.5 & o & \textbf{84} & -4.881 & 4.642 & 55.2 & 78.8 & o & \textbf{166} & 10.288 & -0.889 & 59.3 & 78.9 & o & \textbf{248} & 0.15 & 7.611 & 58.3 & 76.6 & o\\
\textbf{3} & -3.248 & -3.854 & 55.3 & 79.6 & o & \textbf{85} & -4.319 & -5.765 & 54.3 & 79.7 & o & \textbf{167} & 90.004 & -1.159 & 38.0 & 75.7 & s & \textbf{249} & 1.633 & 8.442 & 56.9 & 73.6 & o\\
\textbf{4} & -96.903 & 0.164 & 53.7 & 78.6 & s & \textbf{86} & -4.31 & 5.527 & 58.0 & 77.0 & o & \textbf{168} & 8.142 & -2.881 & 59.5 & 80.6 & o & \textbf{250} & 94.296 & -1.204 & 37.6 & 76.9 & s\\
\textbf{5} & -1.602 & -2.836 & 56.3 & 80.7 & o & \textbf{87} & -3.677 & -6.642 & 57.5 & 84.4 & o & \textbf{169} & 9.243 & -0.81 & 62.0 & 79.6 & o & \textbf{251} & 4.332 & 3.765 & 60.5 & 79.4 & o\\
\textbf{6} & -0.527 & 2.739 & 55.4 & 77.6 & o & \textbf{88} & -4.805 & -6.726 & 54.9 & 80.7 & o & \textbf{170} & 7.611 & 3.889 & 60.8 & 78.1 & o & \textbf{252} & 4.85 & 2.827 & 60.4 & 79.6 & o\\
\textbf{7} & -1.619 & -4.792 & 40.3 & 60.6 & o & \textbf{89} & -8.075 & -1.009 & 53.9 & 80.2 & o & \textbf{171} & 7.0 & -6.615 & 55.6 & 80.1 & o & \textbf{253} & 2.23 & 5.584 & 55.3 & 76.2 & o\\
\textbf{8} & -3.77 & -0.969 & 54.4 & 78.9 & o & \textbf{90} & -5.991 & 4.642 & 53.8 & 78.7 & o & \textbf{172} & 7.633 & -3.81 & 62.1 & 80.6 & o & \textbf{254} & 2.204 & -7.575 & 56.9 & 78.6 & o\\
\textbf{9} & -1.044 & 3.726 & 56.4 & 76.5 & o & \textbf{91} & -90.801 & 0.177 & 45.2 & 79.5 & s & \textbf{173} & 5.42 & -7.535 & 59.8 & 81.3 & o & \textbf{255} & 1.146 & 5.664 & 57.1 & 77.1 & o\\
\textbf{10} & -3.279 & 0.004 & 56.9 & 80.5 & o & \textbf{92} & -4.854 & -4.801 & 53.7 & 79.1 & o & \textbf{174} & 8.146 & -4.748 & 61.2 & 81.6 & o & \textbf{256} & 0.558 & 6.562 & 60.6 & 79.2 & o\\
\textbf{11} & -1.071 & -3.854 & 51.8 & 79.5 & o & \textbf{93} & -5.385 & -3.867 & 55.7 & 80.7 & o & \textbf{175} & 9.212 & -2.881 & 61.1 & 82.1 & o & \textbf{257} & 0.597 & -6.646 & 61.6 & 84.0 & o\\
\textbf{12} & -2.111 & 3.726 & 56.4 & 78.1 & o & \textbf{94} & -6.531 & 1.779 & 54.0 & 79.7 & o & \textbf{176} & 8.642 & -1.867 & 59.4 & 81.0 & o & \textbf{258} & 3.894 & 4.717 & 61.3 & 81.1 & o\\
\textbf{13} & -2.142 & -1.907 & 56.0 & 80.6 & o & \textbf{95} & -5.912 & -2.805 & 55.4 & 82.6 & o & \textbf{177} & 8.549 & 1.823 & 61.3 & 79.3 & o & \textbf{259} & 2.814 & 4.841 & 61.2 & 81.6 & o\\
\textbf{14} & -1.637 & -0.969 & 56.8 & 81.5 & o & \textbf{96} & -6.491 & 3.739 & 52.7 & 77.2 & o & \textbf{178} & 4.916 & -8.465 & 57.7 & 80.6 & o & \textbf{260} & 2.681 & -4.792 & 59.2 & 81.2 & o\\
\textbf{15} & -3.73 & -2.885 & 55.7 & 82.3 & o & \textbf{97} & -7.071 & -1.018 & 54.7 & 81.0 & o & \textbf{179} & 5.942 & -6.606 & 59.9 & 80.8 & o & \textbf{261} & 2.137 & 7.681 & 60.6 & 77.8 & o\\
\textbf{16} & -0.54 & 0.889 & 57.8 & 80.6 & o & \textbf{98} & -5.907 & -4.801 & 53.8 & 79.3 & o & \textbf{180} & 8.066 & 2.81 & 61.2 & 79.1 & o & \textbf{262} & 5.314 & 1.819 & 58.1 & 78.3 & o\\
\textbf{17} & -3.257 & -1.903 & 54.5 & 80.1 & o & \textbf{99} & -7.084 & 2.743 & 50.8 & 74.5 & o & \textbf{181} & 6.527 & -7.469 & 61.1 & 81.4 & o & \textbf{263} & 2.735 & 6.677 & 56.4 & 75.8 & o\\
\textbf{18} & -2.124 & 1.827 & 58.2 & 81.9 & o & \textbf{100} & -8.106 & 0.85 & 53.5 & 79.5 & o & \textbf{182} & 91.053 & -1.212 & 39.5 & 76.9 & s & \textbf{264} & 2.235 & -5.757 & 60.4 & 82.1 & o\\
\textbf{19} & -2.146 & -3.854 & 55.0 & 80.0 & o & \textbf{101} & -4.881 & 6.58 & 55.4 & 76.7 & o & \textbf{183} & 8.726 & -3.743 & 58.5 & 80.6 & o & \textbf{265} & 0.04 & 5.575 & 51.4 & 68.3 & o\\
\textbf{20} & 0.0 & 0.0 & 54.5 & 79.7 & o & \textbf{102} & -8.69 & -0.058 & 53.5 & 78.7 & o & \textbf{184} & 6.075 & 8.553 & 58.7 & 75.4 & o & \textbf{266} & 3.872 & -4.792 & 58.3 & 79.9 & o\\
\textbf{21} & -0.496 & -2.836 & 58.0 & 82.1 & o & \textbf{103} & -2.686 & -8.478 & 53.4 & 78.7 & o & \textbf{185} & 7.04 & -4.743 & 61.6 & 81.7 & o & \textbf{267} & 5.434 & 0.049 & 57.9 & 79.5 & o\\
\textbf{22} & -96.08 & 0.181 & 37.8 & 68.5 & s & \textbf{104} & -88.681 & 0.173 & 45.6 & 77.5 & s & \textbf{186} & 9.173 & 2.823 & 63.2 & 79.5 & o & \textbf{268} & 4.938 & -2.876 & 61.1 & 82.8 & o\\
\textbf{23} & -1.637 & 2.73 & 56.8 & 79.7 & o & \textbf{105} & -6.434 & -3.858 & 52.2 & 77.3 & o & \textbf{187} & 7.58 & -5.717 & 60.9 & 83.0 & o & \textbf{269} & 1.084 & -5.765 & 56.4 & 79.9 & o\\
\textbf{24} & -1.075 & 0.004 & 57.9 & 81.9 & o & \textbf{106} & -5.327 & -5.765 & 52.5 & 77.4 & o & \textbf{188} & 5.942 & 6.58 & 60.5 & 78.7 & o & \textbf{270} & 4.903 & -0.929 & 58.7 & 80.3 & o\\
\textbf{25} & -3.796 & 2.739 & 55.5 & 79.7 & o & \textbf{107} & -5.429 & 3.708 & 53.2 & 75.9 & o & \textbf{189} & 6.513 & -5.708 & 62.6 & 82.9 & o & \textbf{271} & 5.42 & -1.867 & 60.2 & 80.6 & o\\
\textbf{26} & -1.695 & 0.885 & 56.9 & 80.2 & o & \textbf{108} & -3.124 & 7.491 & 54.2 & 74.0 & o & \textbf{190} & 9.699 & -1.863 & 61.5 & 81.8 & o & \textbf{272} & 6.588 & -0.013 & 57.9 & 79.2 & o\\
\textbf{27} & -0.504 & -4.783 & 57.1 & 82.9 & o & \textbf{109} & -7.013 & -2.885 & 53.5 & 80.5 & o & \textbf{191} & 7.035 & 4.735 & 60.9 & 79.6 & o & \textbf{273} & 6.009 & -0.876 & 59.9 & 80.7 & o\\
\textbf{28} & -1.084 & -1.912 & 56.0 & 81.3 & o & \textbf{110} & -7.646 & 1.774 & 52.3 & 77.4 & o & \textbf{192} & 8.177 & 4.668 & 62.4 & 80.6 & o & \textbf{274} & 4.912 & 0.942 & 58.2 & 78.3 & o\\
\textbf{29} & -2.721 & -2.845 & 56.7 & 82.9 & o & \textbf{111} & -7.597 & -1.96 & 51.6 & 77.1 & o & \textbf{193} & 5.456 & 7.69 & 59.2 & 78.0 & o & \textbf{275} & 93.235 & -1.159 & 37.8 & 74.0 & s\\
\textbf{30} & -93.973 & 0.181 & 39.6 & 72.0 & s & \textbf{112} & -3.204 & -7.504 & 52.8 & 78.2 & o & \textbf{194} & 90.597 & -0.155 & 38.6 & 76.3 & s & \textbf{276} & 3.301 & -5.761 & 58.9 & 80.6 & o\\
\textbf{31} & -95.044 & 0.186 & 45.5 & 81.3 & s & \textbf{113} & -4.199 & 7.544 & 54.0 & 75.5 & o & \textbf{195} & 9.177 & 0.973 & 62.8 & 81.8 & o & \textbf{277} & 3.323 & 5.584 & 54.0 & 75.4 & o\\
\textbf{32} & -0.473 & 4.801 & 58.7 & 80.2 & o & \textbf{114} & -3.819 & 6.668 & 52.6 & 75.2 & o & \textbf{196} & 8.695 & 3.819 & 63.7 & 80.1 & o & \textbf{278} & 5.978 & 0.934 & 58.3 & 77.3 & o\\
\textbf{33} & -1.655 & 4.73 & 56.4 & 79.4 & o & \textbf{115} & -6.5 & -1.96 & 52.5 & 77.9 & o & \textbf{197} & 10.814 & 0.088 & 58.5 & 81.5 & o & \textbf{279} & 1.668 & -6.646 & 59.2 & 81.5 & o\\
\textbf{34} & -4.35 & -1.85 & 54.2 & 80.5 & o & \textbf{116} & -7.049 & 0.845 & 51.5 & 77.9 & o & \textbf{198} & 7.566 & 5.633 & 55.7 & 76.2 & o & \textbf{280} & 1.659 & -8.469 & 54.5 & 76.6 & o\\
\textbf{35} & -2.739 & 2.774 & 55.6 & 78.5 & o & \textbf{117} & -89.721 & 0.181 & 46.2 & 80.6 & s & \textbf{199} & 5.027 & 8.522 & 58.6 & 75.5 & o & \textbf{281} & 1.097 & -7.571 & 56.0 & 77.7 & o\\
\textbf{36} & -1.062 & 1.827 & 59.8 & 82.2 & o & \textbf{118} & -5.978 & 2.761 & 51.4 & 73.9 & o & \textbf{200} & 9.699 & 1.881 & 61.6 & 81.4 & o & \textbf{282} & 2.743 & -6.642 & 52.7 & 73.0 & o\\
\textbf{37} & -2.133 & 0.004 & 57.9 & 81.5 & o & \textbf{119} & -3.659 & 8.456 & 53.0 & 73.2 & o & \textbf{201} & 10.283 & 1.022 & 58.0 & 82.2 & o & \textbf{283} & 1.062 & 3.779 & 57.6 & 77.4 & o\\
\textbf{38} & -4.385 & 1.788 & 55.8 & 79.6 & o & \textbf{120} & -5.367 & 5.562 & 59.1 & 78.2 & o & \textbf{202} & 6.951 & 6.588 & 57.9 & 76.6 & o & \textbf{284} & 3.274 & 3.752 & 59.0 & 77.5 & o\\
\textbf{39} & -2.752 & 0.894 & 51.3 & 75.1 & o & \textbf{121} & -3.695 & -8.5 & 53.8 & 78.6 & o & \textbf{203} & 3.239 & -7.584 & 56.8 & 80.9 & o & \textbf{285} & 2.19 & 0.044 & 59.0 & 79.3 & o\\
\textbf{40} & -97.681 & 0.279 & 41.8 & 75.6 & s & \textbf{122} & -2.531 & 8.593 & 52.5 & 68.5 & o & \textbf{204} & 7.08 & -2.788 & 63.1 & 83.9 & o & \textbf{286} & 2.226 & -1.951 & 57.5 & 79.0 & o\\
\textbf{41} & -2.721 & -0.965 & 53.9 & 79.3 & o & \textbf{123} & -7.646 & -0.053 & 50.5 & 73.2 & o & \textbf{205} & 6.611 & 3.92 & 56.5 & 75.6 & o & \textbf{287} & 0.615 & -2.929 & 58.8 & 81.3 & o\\
\textbf{42} & -3.332 & 1.783 & 55.4 & 78.8 & o & \textbf{124} & -4.274 & -7.58 & 47.5 & 71.0 & o & \textbf{206} & 5.987 & -4.739 & 59.3 & 80.1 & o & \textbf{288} & 2.743 & -0.925 & 59.3 & 80.2 & o\\
\textbf{43} & -0.509 & -0.965 & 55.4 & 79.6 & o & \textbf{125} & -9.668 & -1.973 & 52.9 & 78.8 & o & \textbf{207} & 3.867 & 8.562 & 59.2 & 75.3 & o & \textbf{289} & 2.217 & 3.774 & 58.0 & 77.2 & o\\
\textbf{44} & -3.265 & 3.739 & 54.5 & 76.5 & o & \textbf{126} & -5.341 & -7.584 & 50.3 & 75.0 & o & \textbf{208} & 5.425 & -5.712 & 57.7 & 79.9 & o & \textbf{290} & 1.646 & 2.823 & 57.9 & 77.6 & o\\
\textbf{45} & -5.416 & -0.044 & 52.8 & 77.6 & o & \textbf{127} & -6.429 & -7.5 & 49.5 & 75.8 & o & \textbf{209} & 8.071 & 0.938 & 61.5 & 81.2 & o & \textbf{291} & 96.956 & -0.252 & 41.7 & 78.0 & s\\
\textbf{46} & -2.009 & 7.597 & 54.7 & 75.5 & o & \textbf{128} & -6.513 & 5.558 & 56.4 & 73.3 & o & \textbf{210} & 4.296 & -5.761 & 61.0 & 82.7 & o & \textbf{292} & -0.009 & -3.85 & 55.7 & 78.7 & o\\
\textbf{47} & -4.358 & 3.743 & 53.5 & 76.4 & o & \textbf{129} & -8.606 & -3.903 & 51.9 & 77.6 & o & \textbf{211} & 6.522 & -3.81 & 59.4 & 80.1 & o & \textbf{293} & 0.606 & 4.624 & 56.8 & 77.3 & o\\
\textbf{48} & -3.252 & 5.544 & 57.6 & 76.9 & o & \textbf{130} & -7.527 & -3.858 & 50.8 & 77.4 & o & \textbf{212} & 7.013 & 0.934 & 59.1 & 79.1 & o & \textbf{294} & 2.699 & 0.934 & 56.0 & 76.9 & o\\
\textbf{49} & -2.708 & 6.704 & 54.7 & 76.0 & o & \textbf{131} & -7.597 & 3.739 & 55.3 & 79.5 & o & \textbf{213} & 8.655 & 0.04 & 61.0 & 80.0 & o & \textbf{295} & 0.566 & -0.969 & 56.4 & 79.7 & o\\
\textbf{50} & -1.633 & -6.69 & 50.5 & 77.4 & o & \textbf{132} & -87.611 & 0.177 & 48.4 & 82.5 & s & \textbf{214} & 5.473 & 5.619 & 52.3 & 75.5 & o & \textbf{296} & 1.659 & -2.934 & 58.0 & 80.5 & o\\
\textbf{51} & -5.425 & -1.956 & 48.2 & 71.6 & o & \textbf{133} & -8.155 & 2.748 & 33.5 & 57.8 & o & \textbf{215} & 6.0 & -2.836 & 61.8 & 83.7 & o & \textbf{297} & 3.801 & 2.819 & 58.4 & 77.5 & o\\
\textbf{52} & -89.252 & -0.739 & 41.2 & 69.6 & s & \textbf{134} & -9.19 & -1.031 & 54.8 & 79.0 & o & \textbf{216} & 7.588 & -1.867 & 61.5 & 82.5 & o & \textbf{298} & 3.27 & 0.049 & 56.5 & 77.7 & o\\
\textbf{53} & -5.407 & 1.792 & 54.7 & 77.4 & o & \textbf{135} & -6.0 & 6.54 & 54.0 & 75.9 & o & \textbf{217} & 5.451 & 3.805 & 60.0 & 79.5 & o & \textbf{299} & 1.615 & -4.792 & 58.0 & 80.7 & o\\
\textbf{54} & -0.903 & 7.513 & 56.9 & 75.6 & o & \textbf{136} & -7.013 & -6.704 & 49.5 & 77.8 & o & \textbf{218} & 6.416 & 1.827 & 60.1 & 79.8 & o & \textbf{300} & 98.049 & -0.204 & 38.2 & 74.7 & s\\
\textbf{55} & -4.858 & 2.748 & 54.3 & 77.7 & o & \textbf{137} & -4.681 & 8.54 & 56.2 & 76.2 & o & \textbf{219} & 7.058 & -0.889 & 62.8 & 83.2 & o & \textbf{301} & 3.832 & -2.925 & 57.9 & 80.0 & o\\
\textbf{56} & -2.686 & -6.646 & 55.4 & 81.3 & o & \textbf{138} & -7.022 & -4.796 & 47.7 & 72.7 & o & \textbf{220} & 5.925 & 2.858 & 59.8 & 79.9 & o & \textbf{302} & 2.199 & -3.81 & 54.7 & 76.8 & o\\
\textbf{57} & -1.031 & -5.761 & 51.0 & 75.5 & o & \textbf{139} & -5.336 & 7.469 & 54.9 & 74.7 & o & \textbf{221} & 4.841 & 6.575 & 59.3 & 78.3 & o & \textbf{303} & 1.677 & -0.92 & 56.6 & 78.6 & o\\
\textbf{58} & -2.664 & -4.788 & 44.3 & 69.7 & o & \textbf{140} & -9.226 & 2.735 & 55.3 & 79.4 & o & \textbf{222} & 7.487 & 1.819 & 60.6 & 79.9 & o & \textbf{304} & 3.845 & -0.929 & 58.6 & 79.4 & o\\
\textbf{59} & -3.739 & -4.788 & 48.2 & 72.7 & o & \textbf{141} & -8.088 & 4.619 & 52.0 & 74.8 & o & \textbf{223} & 2.774 & 8.575 & 58.7 & 74.9 & o & \textbf{305} & 4.327 & 0.044 & 57.8 & 78.3 & o\\
\textbf{60} & -1.447 & 8.628 & 52.3 & 69.2 & o & \textbf{142} & -5.85 & -6.726 & 46.3 & 75.1 & o & \textbf{224} & 4.323 & 7.575 & 59.3 & 77.7 & o & \textbf{306} & 1.066 & -3.863 & 56.2 & 78.6 & o\\
\textbf{61} & -4.31 & -3.858 & 53.5 & 78.5 & o & \textbf{143} & -9.757 & 1.783 & 52.1 & 78.2 & o & \textbf{225} & 8.173 & -0.889 & 61.8 & 82.2 & o & \textbf{307} & 1.102 & 1.823 & 56.9 & 78.3 & o\\
\textbf{62} & -2.102 & -5.761 & 54.5 & 79.4 & o & \textbf{144} & -7.035 & 4.642 & 53.8 & 75.9 & o & \textbf{226} & 6.969 & 2.814 & 58.9 & 77.3 & o & \textbf{308} & 0.571 & 0.889 & 57.4 & 78.9 & o\\
\textbf{63} & -1.049 & -7.58 & 53.8 & 79.1 & o & \textbf{145} & -9.757 & -0.053 & 54.0 & 77.4 & o & \textbf{227} & 6.071 & 4.85 & 59.0 & 78.7 & o & \textbf{309} & 95.938 & -0.252 & 38.1 & 74.3 & s\\
\textbf{64} & -94.345 & -0.735 & 45.7 & 81.2 & s & \textbf{146} & -4.814 & -8.513 & 53.7 & 77.9 & o & \textbf{228} & 6.522 & -1.863 & 59.7 & 80.0 & o & \textbf{310} & 0.018 & 3.73 & 56.2 & 75.7 & o\\
\textbf{65} & -4.876 & 0.841 & 52.2 & 77.4 & o & \textbf{147} & -7.478 & -5.761 & 52.1 & 78.4 & o & \textbf{229} & 4.398 & 5.571 & 55.7 & 77.6 & o & \textbf{311} & 2.704 & 2.823 & 56.2 & 76.6 & o\\
\textbf{66} & -0.522 & -6.69 & 49.5 & 75.0 & o & \textbf{148} & -5.73 & 8.544 & 55.9 & 75.8 & o & \textbf{230} & 3.863 & -8.513 & 54.2 & 76.5 & o & \textbf{312} & 0.562 & -4.788 & 57.4 & 80.6 & o\\
\textbf{67} & -5.987 & 0.845 & 52.6 & 77.6 & o & \textbf{149} & -7.049 & 6.553 & 50.4 & 75.1 & o & \textbf{231} & 4.929 & -6.642 & 59.9 & 80.5 & o & \textbf{313} & 2.08 & 1.77 & 55.0 & 75.9 & o\\
\textbf{68} & -1.606 & 6.695 & 55.3 & 75.9 & o & \textbf{150} & -7.611 & 5.655 & 58.7 & 76.2 & o & \textbf{232} & 4.296 & -7.58 & 58.1 & 80.7 & o & \textbf{314} & 3.261 & -3.858 & 60.1 & 79.7 & o\\
\textbf{69} & 0.022 & -7.58 & 55.8 & 79.9 & o & \textbf{151} & -8.022 & -4.792 & 52.9 & 77.0 & o & \textbf{233} & 7.597 & 0.044 & 59.4 & 79.2 & o & \textbf{315} & 4.323 & -1.907 & 57.8 & 78.6 & o\\
\textbf{70} & -3.81 & 4.646 & 55.0 & 76.6 & o & \textbf{152} & -8.726 & 3.717 & 55.4 & 79.4 & o & \textbf{234} & 3.279 & 7.58 & 60.2 & 77.5 & o & \textbf{316} & 2.726 & -2.929 & 60.6 & 81.6 & o\\
\textbf{71} & -6.482 & -0.04 & 54.5 & 79.6 & o & \textbf{153} & -8.726 & 1.792 & 52.4 & 78.3 & o & \textbf{235} & 3.757 & 6.602 & 57.7 & 75.4 & o & \textbf{317} & 1.712 & 4.832 & 58.7 & 78.7 & o\\
\textbf{72} & -5.956 & -1.013 & 52.3 & 77.8 & o & \textbf{154} & -10.765 & -0.04 & 55.0 & 79.8 & o & \textbf{236} & 2.735 & -8.473 & 55.1 & 76.9 & o & \textbf{318} & 1.093 & -1.907 & 56.2 & 78.4 & o\\
\textbf{73} & 0.031 & -5.677 & 54.9 & 79.1 & o & \textbf{155} & -88.159 & -0.752 & 49.6 & 81.9 & s & \textbf{237} & 91.637 & -0.257 & 36.3 & 73.6 & s & \textbf{319} & 4.208 & 1.774 & 56.5 & 76.8 & o\\
\textbf{74} & -2.142 & 5.549 & 58.6 & 77.2 & o & \textbf{156} & -9.071 & -2.92 & 53.5 & 80.4 & o & \textbf{238} & 3.823 & -6.65 & 59.3 & 81.1 & o & \textbf{320} & 1.137 & 0.049 & 56.3 & 78.6 & o\\
\textbf{75} & -4.841 & -2.934 & 54.4 & 79.8 & o & \textbf{157} & -7.969 & -2.796 & 54.9 & 81.0 & o & \textbf{239} & 92.659 & -0.252 & 35.9 & 74.4 & s & \textbf{321} & 3.265 & -1.907 & 55.8 & 77.5 & o\\
\textbf{76} & -0.681 & 6.677 & 57.5 & 77.2 & o & \textbf{158} & -10.195 & -1.022 & 54.5 & 78.7 & o & \textbf{240} & 5.416 & -3.819 & 59.5 & 80.1 & o & \textbf{322} & 0.504 & 2.708 & 55.5 & 75.8 & o\\
\textbf{77} & -91.425 & -0.712 & 47.1 & 84.4 & s & \textbf{159} & -9.159 & 0.845 & 48.4 & 73.3 & o & \textbf{241} & 4.965 & -4.655 & 60.3 & 81.7 & o & \textbf{323} & 0.0 & 1.832 & 59.3 & 79.4 & o\\
\textbf{78} & -1.624 & -8.504 & 52.6 & 77.2 & o & \textbf{160} & -8.646 & -1.96 & 52.2 & 76.5 & o & \textbf{242} & 93.752 & -0.257 & 40.2 & 77.4 & s & \textbf{324} & 1.633 & 0.889 & 55.0 & 76.8 & o\\
\textbf{79} & -2.73 & 4.659 & 55.8 & 77.1 & o & \textbf{161} & -6.442 & -5.77 & 53.1 & 77.4 & o & \textbf{243} & 1.659 & 6.686 & 60.5 & 80.1 & o & \textbf{325} & 3.85 & 0.938 & 57.4 & 76.9 & o\\
\textbf{80} & -3.204 & -5.765 & 56.0 & 79.7 & o & \textbf{162} & -5.854 & -8.509 & 0.8 & 6.8 & o & \textbf{244} & 4.358 & -3.81 & 59.9 & 81.6 & o & \textbf{326} & -0.013 & -1.903 & 56.9 & 77.5 & o\\
\textbf{81} & -2.106 & -7.562 & 54.8 & 80.0 & o & \textbf{163} & -6.487 & 7.588 & 49.5 & 74.4 & o & \textbf{245} & 0.606 & -8.509 & 55.3 & 78.3 & o & \textbf{327} & 3.15 & 1.77 & 56.3 & 75.2 & o\\
\textbf{82} & -4.841 & -1.018 & 55.8 & 80.7 & o & \textbf{164} & -10.23 & 0.85 & 50.6 & 72.7 & o & \textbf{246} & 0.677 & 8.566 & 56.4 & 72.6 & o\\ \bottomrule
\enddata
\end{deluxetable}

\end{document}